\documentclass[onecolumn,preprintnumbers,amsmath,amssymb,nofootinbib,12pt]{revtex4}
\usepackage{graphicx}
\usepackage{dcolumn}
\usepackage{bm}

\input psfig.sty

\newcommand{\bq}{\begin{equation}}
\newcommand{\eq}{\end{equation}}
\newcommand{\bqn}{\begin{eqnarray}}
\newcommand{\eqn}{\end{eqnarray}}

\newcommand{\lb}{\label}

\def\gappr{\lower 3pt\hbox{$\buildrel > \over \sim\;$}}
\def\gappl{\lower 3pt\hbox{$\buildrel < \over \sim\;$}}
\def\limiter{\lower 7pt\hbox{$\buildrel{\textstyle\longrightarrow}\over{\scriptscriptstyle ~~s\rightarrow\infty~~}\;$}}
\def\ablim{\lower 9pt\hbox{$\buildrel{\textstyle\longrightarrow}\over{\scriptscriptstyle ~~~a\rightarrow b~~~}\;$}}
\def\x0lim{\lower 11pt\hbox{$\buildrel{\textstyle\longrightarrow}\over{\scriptscriptstyle ~~x^0\rightarrow-\infty~~}\;$}}
\def\xlim{\lower 8.5pt\hbox{$\buildrel{\textstyle\longrightarrow}\over{\scriptscriptstyle ~~x_\pm\rightarrow-\infty~~}\;$}}
\def\T0lim{\lower 11pt\hbox{$\buildrel{\textstyle\longrightarrow}\over{\scriptscriptstyle ~~T\rightarrow0~~}\;$}}
\def\Tlim{\lower 8.5pt\hbox{$\buildrel{\textstyle\longrightarrow}\over{\scriptscriptstyle~~T\rightarrow\infty~~}\;$}}
\def\Tmg1{\lower 8.5pt\hbox{$\buildrel{\textstyle\longrightarrow}\over{\scriptscriptstyle~~T>>1~~}\;$}}
\def\pdot{\raise 1.5pt\hbox{.}}

\def\dal{\hbox{$\sqcup$\hbox to 0pt{\hss$\sqcap$}}}

         \def\ln{{\rm ln}}
         \def\ex{{\rm e}}
         \def\dal{\hbox{$\sqcup$\hbox to 0pt{\hss$\sqcap$}}}

\begin{document}
            
\title{\Large On the anomalous mass defect of strange stars \\ in the Field Correlator Method}
\author{F. I. M. Pereira}\email{flavio@on.br}
\affiliation{Observat\'orio Nacional, MCTI, Rua Gal. Jos\'e Cristino 77, 
20921-400 Rio de Janeiro RJ, Brazil}
\date{\today}

\begin{abstract}
 We investigate general aspects of the mass defects of strange stars in the context of the Field 
Correlator Method, without magnetic field. 
 The main parameters of the model that enter the corresponding nonperturbative 
equation of state of the quark gluon plasma are the gluon condensate $G_2$ and 
the large distance static $Q{\bar Q}$ potential $V_1$.

 We calculate mass defects of stellar configurations in the central density range 
$11<\log\rho_c<18$. 
 In general, the mass defects are strongly dependent on the model parameters.  
 For a large range of values of $G_2$ and $V_1$, we obtain anomalous mass defects 
with magnitudes around $10^{53}\,$erg\,, of the same order of the observed 
energies of gamma-ray bursts and neutrino emissions in SN1987A, and of the theoretically 
predicted energies of the quark-novae explosions. 

\end{abstract}

\maketitle

 
\section{\bf Introduction}
\lb{int}

 In pioneer works, V. L. Ambartsumyan and G. S. Saakyan considered the question of 
superdense stellar matter made of a degenerate gas of elementary particles, comprising 
neutron, protons, hyperons and electrons, at zero temperature\,\cite{Amb1,Amb2}. 
 Investigations of internal structures of these compact configurations led to the 
possibility of stellar transitions of explosive character, from a metastable state to a stable 
state, with great amounts of liberated energy\,\cite{Amb3,Sak}. 
 These transitions were related to stars with negative (or anomalous) mass defects 
characterized by energy excesses with respect to the energies they would have to be (stable) 
bound systems. 
  Anomalous mass defects were interpreted in terms of a catastrophic additivity violation 
of the internal energy due to the very intense gravitational fields in the interior of such 
superdense stellar configurations\,\cite{Amb4,Amb5}. 
  An important aspect of these early works rests on the fact that they included the strange 
baryons - hyperons - in the degenerate (strange) nuclear plasma together with neutrons, 
protons and electrons. 

 The baryons being made of quarks, it appeared to be natural to expect unbound quarks to exist 
in the interior of hyperdense stars.
 Within this assumption, and in an epoch when the physics of the strong interactions was very  
incomplete, N. Itoh considered the possibility of hypothetical compact stars made of pure quark 
matter\,\cite{Ito}.  

 With the subsequent developments of the strong interactions theory, new interests came into 
play connecting the strange nuclear plasma with the physics of quarks.
 The nuclear interactions within the superdense stellar matter turned out to be described in 
terms of the baryon constituent quarks. 
 In this context, the strange quark matter (SQM) concept appeared leading to the conjecture of 
the absolute stability of nuclear matter. 

 The Bodmer-Terazawa-Witten conjecture\,\cite{Bod,Wit}, which says that the SQM should be the true 
ground state of the nuclear matter, has attracted a great deal of attention.
 SQM is a type of quark matter made of approximately equal amounts of u, d and s quarks with 
a small admixture of electrons in order to maintain the charge neutrality. 
 Its energy per baryon might be lower than the one in ordinary nuclear matter.
 The SQM properties are of great importance for nuclear physics and astrophysics.
 In the nuclear physics context, E. Farhi and R. L. Jaffe  studied, within the MIT Bag Model 
(BM), the dependence of the SQM stability on the model parameters, namely, the bag constant 
$B$, the strong interactions coupling constant $\alpha_c$, and the strange quark mass 
$m_{\rm s}$\,\cite{Far}.
 In the astrophysical context, the Bodmer-Terazawa-Witten conjecture has proved to be of great 
significance for the physics of the strange stars. 

 Since the 1980's, the properties of strange stars have been considered within the BM\,
\cite{AFO,HZS,Var2,Koh}. 
 In the BM, the quarks enter the equation of state (EOS) as free particles, with the quark 
confinement being represented by the bag constant $B$. 
 Another well known model is the Nambu-Jona-Lasinio model\,\cite{NJL1,NJL2} used to investigate 
quark matter properties in compact stars in Refs.\,\cite{DPM1,DPM2}. 
 It exhibits chiral symmetry breaking, but the quark confinement is not explicitly included. 
 In alternative investigations, mass density dependent models were considered to represent 
confinement in Refs.\,\cite{BeL,LuB,Day1,Day2}.
 In a recent work, the Richardson potential\,\cite{Ric}, which incorporates the asymptotic 
freedom and linear quark confinement, and also used in Refs.\,\cite{Day1,Day2}, was considered 
to investigate the SQM in strong magnetic field by the authors of Ref.\,\cite{SHS}.
 
 Due to the highly nonlinear character of the theory of strong interactions, it was difficult 
to deal with a definite model of EOS naturally including the quark confinement in terms of the 
interactions between quarks and antiquarks, and gluons. 
 Thanks to the developments of quantum chromodynamics (QCD), the fundamental theory of the 
strong interactions, great advances have been made to derive an EOS, including 
perturbative and/or nonperturbative effects of confinement, to describe quark matter at all 
finite densities and temperatures. 

 Recently, Yu. A. Simonov derived, from the first principles, the nonperturbative equation of 
state (NPEOS) of the quark-gluon plasma in the framework of the Field Correlator Method (FCM) 
\cite{Si6}.
 In the FCM (for a review see Ref.\,\cite{DiG} and references therein), the dynamics of 
confinement is naturally included in terms of the color-electric and color-magnetic 
correlators.
 The main parameters of the model that enter the NPEOS are the gluon condensate $G_2$ and the 
large distance static ${\rm Q\bar Q}$ potential $V_1$, at fixed quark masses and temperature.
 The model covers the entire phase diagram plane, from the low $T$ and large $\mu$ regime to 
the large $T$ and low $\mu$ regime. 
 By connecting the FCM and lattice simulations, at $\mu_c=0$\,, the critical temperature turned 
out to be $T_c\sim170$ MeV for $G_2=0.00682\;{\rm GeV}^4$\,\cite{ST1,ST2}. 
 Very recently, V. D. Orlovsky and Yu. A. Simonov considered the quark-hadron thermodynamics in 
the presence of the magnetic field within the FCM\,\cite{OS}. 
  
 Astrophysical applications of the FCM have been made in the study of neutron stars interiors 
\cite{Bur,Bal,Plu} and in the early Universe cosmology\,\cite{Cas}. 
 The authors of Ref.\,\cite{Bom,Log1,Log2} also applied the FCM to the study of phase transitions 
in neutron stars matter and to the investigation of the structural properties and stability of 
hybrid stars.

  Recently we applied the FCM to investigate the properties of the strange stars and the SQM 
stability in Refs.\,\cite{Fla1,Fla2}. 
 Of particular significance is the gradual decrease of the widths of the SQM stability windows 
with the increase of $V_1$\,, being zero at $V_1=0.5\,{\rm GeV}$\,, the value of $V_1$ 
determined from lattice calculations\,\cite{KaZ}. 
 This aspect is of great importance to investigate the existence of strangelets, mainly in the 
case of exploding stars with the liberation of matter/energy into the free space, which we here 
briefly consider (at the end of the present paper).
   
 In the present work, we study the general aspects of the mass defects of strange 
stars, without crust and magnetic field, within the framework of the FCM. 
 We do not consider the crust here by the same reasons we disregarded it in our previous paper 
 \cite{Fla1}.
 Crust contributions to the masses of strange stars have been estimated to be 
$M_{\rm cr}\simeq2.5\times10^{-5}\,M_\odot$\,\cite{AFO}, 
$M_{\rm cr}\simeq(0.9\,-\,2.3)\times10^{-5}\,M_\odot$\,\cite{Var4,Var5}, and 
$M_{\rm cr}\simeq3.4\times10^{-6}\,M_\odot$\,\cite{TLu}.
 As we shall see below, the mass defects magnitudes we obtained in the present work are of 
the order of $10^{53}\,{\rm erg}\simeq0.056\,M_\odot$\,, corresponding to  
$\sim2.2\times10^3\,M_{\rm cr}$\,, $\sim(2.4\,-\,2.6)\times10^3\,M_{\rm cr}$\,, and 
$\sim1.64\times10^4\,M_{\rm cr}$\,, respectively. 
 Crusts are important to investigate compact stars glitches, but we assume here 
that they are insignificant for the purposes of the present work. 
 On the other hand, in this first attempt to investigate the mass defects of the strange 
stars, we are interested in the solutions not affected by preferred directions due to 
magnetic fields, of particular importance to investigate magnetars and soft gamma-ray 
repeaters\,\cite{Hur}  as well.

 Differently from our strategy adopted in Ref. \cite{Fla1}, we here also consider unstable 
solutions of the hydrostatic equilibrium equations of Tolman-Oppenheimer-Volkov. 
 As a result, depending on the values of the model parameters, $G_2$ and $V_1$\,, solutions with 
non-negative and/or negative mass defects along a given sequence of stellar configurations are 
possible. 
 Our aim is to understand the effects of the nonperturbative dynamics of confinement on the binding 
energies of the strange stars.
 We give special attention to the anomalous mass defects and, at the end of the paper, briefly 
comment the corresponding consequences for astrophysical phenomena, such as gamma-ray bursts, 
supernovae neutrinos or quark-novae explosions.

 The present paper is organized as follows. 
 In Sec.\,\ref{npeos} we show the main equations to be used in our calculation.
 In Sec.\,\ref{stc} we present the equations to calculate the important quantities of the stellar 
configurations.
 In Sec.\,\ref{res} we show the results and in Sec. \ref{frmks} we give the final remarks. 

\section{The NPEOS at zero temperature}
\lb{npeos}

 In previous works, we outlined the main features of the FCM and showed the equations 
used to investigate strange stars and strange quark matter properties\,\cite{Fla1,Fla2}.
 So, we now write only the main equations we need here.
 
 For constant $V_1$, the pressure, energy density and number density of a (one flavor) 
quark gas at $T=0$ are given by

\bq
p_q^{SLA}=\frac{N_c}{3\pi^2}\Bigg\{\frac{k_q^3}{4}\sqrt{k_q^2+m_q^2}-
\frac{3}{8}\;m_q^2\bigg[k_q\sqrt{k_q^2+m_q^2}-m_q^2\;\ln\bigg(\frac{k_q+\sqrt{k_q^2+m_q^2}}{m_q}\bigg)
\bigg]\Bigg\}\;,
\lb{pqT0}
\eq

\bqn
\varepsilon_q^{SLA}&=&\frac{N_c}{\pi^2}\Bigg\{\frac{k_q^3}{4}\sqrt{k_q^2+m_q^2}+
\frac{m_q^2}{8}\;\bigg[k_q\sqrt{k_q^2+m_q^2}-m_q^2\;\ln\bigg(\frac{k_q+\sqrt{k_q^2+m_q^2}}{m_q}
\bigg)\bigg]\nonumber\\
&+&\frac{V_1}{2}\;\frac{k_q^3}{3}
\Bigg\}
\lb{eqT0}
\eqn
and 
\bq
n_q^{SLA}=\frac{N_c}{\pi^2}\;\frac{k_q^3}{3}\;,
\lb{nqT0}
\eq
where
\bq
k_q=\sqrt{(\mu_q-V_1/2)^2-m_q^2}\;,\;\;\;\;(q=\rm{u,d,s})\;,
\lb{kq}
\eq 
and $N_c=3$ is the color number; and $SLA $ indicates the single line approximation considered 
in Ref.\,\cite{Si6}.
When $V_1=0$, the ordinary Fermi momentum $k_F$ of a free quark gas is recovered in Eq.\,(\ref{kq}).
 The additional term $(V_1/2)k_q^3/3\;$ in Eq.(\ref{eqT0}) comes from the large distance  
static $Q{\bar Q}$ potential $V_1$.

 Inside a strange star, the weak interaction reactions $d\rightarrow u+\ex+{\bar\nu}_\ex$\,, 
$\ex+u\rightarrow d+\nu_\ex$\,, and $s\rightarrow u+\ex+{\bar\nu}_\ex$\,, 
$u+\ex\rightarrow s+\nu_\ex$\, imply weak equilibrium between quarks, whereas neutrinos and 
anti-neutrinos  leave the star without interaction and their chemical potentials can be set to zero.
 In this case, the chemical equilibrium is given by 
\bq
\mu_d=\mu_u+\mu_\ex\;\;\;{\rm and}\;\;\;\mu_s=\mu_d\,.
\lb{mud}
\eq
 The overall charge neutrality requires that
\bq
\frac{1}{3}(2n^{SLA}_u-n^{SLA}_d-n^{SLA}_s)-n_\ex=0\,,
\lb{chn}
\eq
where $n_i$ is the number density of the particle $i=$u,d,s,e.

 To calculate stellar configurations, with charge neutrality and chemical equilibrium, 
the total pressure and energy density, including electrons are given by
\bq
p=\sum_{q=u,d,s}p^{SLA}_{q}-\Delta|\varepsilon_{\rm vac}|+p_{\rm e}\;,
\lb{pqgl}
\eq
\bq
\varepsilon=\sum_{q=u,d,s}\varepsilon^{SLA}_{q}+
\Delta|\varepsilon_{\rm vac}|+\varepsilon_{\rm e}\;,
\lb{eqgl}
\eq
where 
\bq
\Delta|\varepsilon_{\rm vac}|=\frac{11-\frac{2}{3}N_f}{32}\Delta G_2\;,
\lb{dvac}
\eq
is the vacuum energy density difference between confined and deconfined phases and 
$N_f$ is the number of flavors. 
 The difference between the values of the gluon condensate, as predicted by lattice 
calculations, is $\Delta G_2=G_2(T<T_c)-G_2(T>T_c)\simeq \frac{1}{2}G_2$ \cite{ST1,ST2}.

 In order to obtain the numerical correspondence between FCM and BM, we make the 
identifications: $\Delta|\varepsilon_{\rm vac}|=B$ and $V_1=0$. 
 However, we here emphasize that $\Delta|\varepsilon_{\rm vac}|$ is essentially a 
nonperturbative quantity. 
 For the quark masses, we use $m_u=5$ MeV, $m_d=7$ MeV and $m_s=150$ MeV.
 The corresponding equations for the degenerate electron gas are similar to the ones above 
and can be easily obtained by making the changes: $N_c\rightarrow1$, $V_1\rightarrow0$, 
$\mu_q\rightarrow\mu_\ex$ and $m_q\rightarrow m_\ex$. 
 We use the same numerical strategy adopted in Ref.\,\cite{Fla1} to calculate strange 
stars configurations.

\section{compact stars configurations} 
\lb{stc}

 Compact stars configurations are calculated by numerical integration of the 
Tolman-Oppenheimer-Volkov hydrostatic equilibrium equations  \cite{ShT,Gle,ZeN}.
 Of particular importance here is the total gravitational mass of a compact star, 
\bq
M=4\pi\int^R_0\varepsilon(r)\,r^2dr\;,
\lb{stm}
\eq
which is the mass that governs the Keplerian orbital motion of the distant gravitating 
bodies around it, as measured by external observers. 
 The proper mass is given by 
\bq
M_P=\int^R_0\varepsilon(r)\,dV(r)\;,
\lb{ma}
\eq
where $dV(r)=4\pi[1-2Gm(r)/r]^{-1/2}\,r^2\,dr$ and $m(r)$ is the mass within a sphere of radius $r$. 
 The proper mass is the sum of the mass elements $dm(r)=\varepsilon(r)\,dV(r)$ measured by a local 
observer.
 The baryonic mass (also called rest mass) of a star is $M_A=N_Am_A$, where 
\bq
N_A=\int^R_0n_A(r)\,dV(r)
\lb{NA}
\eq 
is the number of baryons within the star, $m_A$ is the mass of the baryonic specie $A$, and 
\bq
n_A=\frac{1}{3}(n^{SLA}_u+n^{SLA}_d+n^{SLA}_s)
\lb{na}
\eq
is the baryon number density.
 The baryonic mass has a simple interpretation: it is the mass that the star would have if 
its baryon content were dispersed at infinity.
 In the case of the strange stars (because of the quark confinement), $N_A$ is the equivalent 
number of baryons (not quarks). 
 There is some freedom to choose the baryonic mass. 
 In earlier texts, the baryonic mass was taken as the mass $m_{\rm H}$ of the hydrogen atom\,
\cite{Amb3}; the $^{56}{\rm F_e}$ mass per baryon $m_0\equiv m(^{56}{\rm Fe})/56$\,
\cite{Var2,HTW,ZeN,Var1,Var3}; or the neutron mass $m_n$\,\cite{Sak,Amb4,Fla1,Gle}. 
 We here assume $m_A=m_n$, as in Ref.\,\cite{Fla1}. 
 Comparison of some results with respect to $m_0$ is also made below.
  
 Let us now consider the mass defect of a compact stars we are concerned in the present work. 
 The {\it incomplete mass defect} or, for short, the {\it mass defect} is the difference 
 $\Delta_2M=M_A-M$ (which in our notation\footnote{We here follow the notation according 
to Refs.\,\cite{Amb1,Amb2,Sak,Var1,Var3}.
} is minus the binding energy $E_b$ defined in Refs.\,
\cite{ZeN,Gle}). 
 It corresponds to the energy released to aggregate from infinity the dispersed baryonic matter.
 A stellar configuration is stable if $\Delta_2M>0$ (normal mass defect) and unstable if 
$\Delta_2M<0$ (anomalous mass defect). 

\section{Results}
\lb{res}

 We calculated sequences of strange star configurations for central densities in the range 
$11<\log\rho_c<18$. 
 In general, the forms of the sequences and the respective mass defects of stellar 
configurations strongly depend on the values of the model parameters. 
 A typical example is shown in Fig.\,\ref{mg2v11}, for $G_2=0.006\,\text{GeV}^4\,$\cite{Si6} 
and $V_1=0$. 
 The stellar sequence present three branches delimited at the labeled points 1 and 2, where 
the solid and dashed curves cross itself, and in which $M=M_A$, as shown in panel (a). 
 In the intermediate branch we have $M<M_A$, required by the stability conditions 
against transition to diffuse matter. 
 We have $M>M_A$ at densities $\sim10^{15}{\rm g\,cm^{-3}}$ in the first branch\footnote{Not 
well visible in the scales of panels (a) and (b), but visible in panel (c).
}, and $>5.5\times10^{16}{\rm g\,cm^{-3}}$ in the third branch. 

 An investigation of strange stars within BM, for the values of the bag constant in the range 
$50\,{\rm MeV\,fm^{-3}}\leq B\leq 70\,{\rm MeV\,fm^{-3}}$, showed the absence of the anomalous 
mass defect in strange stars \cite{Var3}.
 Anomalous mass defects does not have been obtained because of the low used values of $B$, which 
numerically correspond (in our calculation with $V_1=0$) to lower values of 
$\Delta|\varepsilon_{\rm vac.}|$ (or $G_2\,$).
  In fact, in the FCM, for $V_1=0$ and $G_2\gappl0.0043\,{\rm GeV}^4$, the stellar configurations 
present normal mass defects. 
 To obtain, within the BM, stellar configurations with anomalous mass defects in the first branch  
we need $B>78.7\,{\rm MeV\,fm^{-3}}$ (corresponding to $G_2>0.0043\,\text{GeV}^4\,$).  
 To obtain anomalous mass defects both in first and in third branches, we need 
 $B\gappr110\,{\rm MeV\,fm^{-3}}$ (corresponding to $G_2\gappr0.006\,\text{GeV}^4\,$). 
 Values of $B$ between $150\,{\rm MeV\,fm^{-3}}$ and $170\,{\rm MeV\,fm^{-3}}$ were considered   
to explain the time elapsed between the transition from a metastable neutron star generated by a 
supernova explosion and the new collapse generating the delayed gamma-ray burst by the authors 
of Ref.\,\cite{Ber}. 
 Moreover, higher values of $B$ up to $337\,{\rm MeV\,fm^{-3}}$ and $353\,{\rm MeV\,fm^{-3}}$  
were considered, but to calculate at nonzero temperatures the quark deconfinement in the cores 
of protoneutron stars\,\cite{Lug}. 

 Of particular interest is the dependence of $M$ with the number of baryons $N_A$  
shown in panel (b), with the labels 1 and 2 as in panel (a). 
  The cusp is at the maximum value of $M$, where $N_A$ is also maximum\footnote{See footnote 2 
in Ref.\,\cite{Fla1}.
}. 
  Also shown is the $M_A$ plot with its upper ``endpoint''\footnote{In reality, it is not an 
endpoint because, at the upper point, the plot comes back along the same straight line.
} at the maximum $M_A$\,. 
 In the upper part of the $M\,{\rm vs.}\,N_A$ plot (above 1) the situation is analogous to that of neutron 
stars in that $dM/dN_A<m_A$ everywhere on the corresponding plot segments; $M<M_A$ in the second 
branch and $M>M_A$ in the third branch\,\cite{ZeN}.
 However, a fact that was not observed in earlier works (because of the EOS used) is that $M>M_A$ 
in the first branch (below 1). 
 Moreover, the slope starts with $dM/dN_A>m_A$\,, turns to $dM/dN_A=m_A$ at an intermediate point  
and then to $dM/dN_A<m_A$ as $N_A$ grows.
 In contrast with neutron star configurations, we have here a situation with $M>M_A$ and 
$dM/dN_A>m_A$ apparently not obeying the $dM/dN_A=m_A\,[1-2GM/R]^{1/2}$ prescription\,\cite{HTW}. 
 This is a characteristic feature of the anomalous mass defects occurring in the first branch 
making evident the role of the confinement effects. 
 
 Panel\,(c) shows the mass defect as function of $M/M_\odot$ with the delimiters 
1 and 2 as in panels (a) and (b). 
 Differently from Refs.\,\cite{Amb1,Amb2,Sak}, the mass defects are also negative in the first 
and third branches\footnote{From now on, we do not mention the value of $\Delta_2M$ at the 
origin because it is obviously zero as $M$\,, $ M_A$ and $N_A\rightarrow0$\,, unless stated 
otherwise.
}. 
 For the given values of the parameters $G_2$ and $V_1$, the maximum $\Delta_2M$ magnitude 
is of the order of $\sim0.15\times10^{53}\,{\rm erg}$ at $M\sim0.26\,M_\odot$ in the first branch, 
and $\sim0.3\times10^{53}\,{\rm erg}$ at the endpoint of the third branch at $M\sim0.9\,M_\odot$.

 A variety of behaviors can be obtained by the variation of the model parameters. 
 Some typical examples are shown in Fig.\,\ref{mg2v12}. 
 Panels (a) and (b) show the case $M<M_A$ for which the mass defect has the normal sign 
($\Delta_2M>0$) everywhere along the sequence of stellar configurations, as in panel (c). 
 Panels (d) and (e) correspond to the limit $M=M_A$ at the maximum mass, but with anomalous 
mass defects at all the other points of the stellar sequence with $M>M_A$, as depicted in 
panel (f). 
 Finally, panels (g) and (h) show the case $M>M_A$, so $\Delta_2M<0$ at all points along 
the sequence, as in panel (i). 
 Notice the pronounced confinement effects on the stellar sequences in panels (d)-(f) and 
(g)-(i).
 This general overview shows us that anomalous mass defects can (in principle) be obtained 
for arbitrary values of the model parameters. 
 If $G_2$ is low, $V_1$ must be increased in order to yield anomalous mass defect; if $V_1$ 
is low, $G_2$ must grow in order to produce the same effect. 
 As it was stated above, new results emerge with the use of the NPEOS provided by the FCM: for 
$V_1$ in the range $0\leq V_1\leq0.5\,{\rm GeV}$, stellar configurations with anomalous mass 
defects are possible not only in the first branch but also in the third branch, 
at densities larger than the nuclear one.
 Merely illustratively, we also show the proper mass $M_P$ which is greater than both $M$ and 
$M_A$.

 On account of the above features, in the energy range we are considering, the $G_2-V_1$ 
plane can be divided in three different regions according to the signs of 
$\Delta_2M$ as shown in Fig.\;\ref{d2mg2v1}. 
 In doing so, we obtain three regions. 
 The first region, A, with $\Delta_2M>0$ everywhere on the sequence. 
 In the second region, B, with features analogous to those in Fig.\;\ref{mg2v11}, 
both normal and anomalous mass defects are present in the same stellar sequence. 
 Finally, the third region, C, with $\Delta_2M<0$ along all the sequence of stellar 
configurations.
 The idea of Fig.\;\ref{d2mg2v1} serves to predict values of $V_1$ and $G_2$ according 
to the types of the stellar configurations and the respective mass defects we want. 

 A star with anomalous mass defect has an exceeding stored energy with respect to the   
one needed to form a compact stable bound system.
 In principle, in a given sequence of stellar configurations, any star with $\Delta_2M<0$ 
might explode or implode (for example, in the presence of certain perturbations) with a 
liberation of an enormous amount of energy. 
 In the case of explosion, the scattered matter will have a nonzero kinetic energy at 
infinity.
 
 On the other hand, it is well known that stellar configurations obtained from 
Tolman-Oppenheimer-Volkov equations are stable if $dM/d\rho_c>0$\,, corresponding to the 
stars in the ascending branch of the stellar sequence, as in panel (a) of Fig.\,\ref{mg2v11}. 
 In the descending branch, where $dM/d\rho_c<0$, the stellar configurations are unstable 
against gravitational collapse to black hole. 
 Then, a stable configuration pass from stability to instability at the peak of the sequence 
where $M$, $M_A$ and $N_A$ attain their maximums (for a detailed analysis of stability, 
see Refs.:\,\cite{Gle,HTW}).
 Hence, the investigation of the mass defects in this transition limit may be of particular 
interest.
 
 Among the many possibilities, as the ones shown in Figs.\;\ref{mg2v11} and \ref{mg2v12}, 
let us now consider (both normal and anomalous) mass defects at the maximum masses of the 
stellar configurations, as depicted in Fig.\,\ref{d2mg2v11}. 
 The plots cover a large area in the $G_2-\Delta_2M$ plane, as shown in panel (a).
  It is evident that nonnegative values of $\Delta_2M$ occur for $G_2\gappl0.00927\,{\rm GeV}^4$\;, 
which gives the vacuum energy density 
$\Delta|\varepsilon_{\rm vac}|\gappl0.0013\,{\rm GeV}^4\simeq170\,{\rm MeV\,fm^{-3}}$. 
 Above these values of $G_2$ (or $\Delta|\varepsilon_{\rm vac}|$), $\Delta_2M$ is anomalous  
whatever the values of $V_1$ may be between zero and 0.5\,GeV.
 
 At $V_1=0.5\,{\rm GeV}\,$, the anomalous mass defect attains its maximum magnitude at 
$\simeq6.27\times10^{53}\,{\rm erg}$\,, but for $G_2\simeq0.000625\,{\rm GeV}^4$\,,  
corresponding to a strange star with $M\simeq1.59\,M_\odot$ and $M_A=1.24\,M_\odot$\,.
 In the FCM framework, this is the maximum allowed energy to be liberated in a possible 
explosion.
 Such a star has a fraction  of stored energy around $|\Delta_2M|/M\sim22\,\%\,$, but it is 
not a maximum.
 For instance, in the range $0\leq G_2\gappl0.0095\,{\rm GeV}^4$\,, the fractions of the mass 
excess may be as large as 
$\sim25\,\%$\, for $V_1=0.3\,{\rm GeV}$ and $G_2=0.006\,{\rm GeV}^4$\,; 
$\sim34\,\%$\, for $V_1=0.4\,{\rm GeV}$\, and $G_2=0.007\,{\rm GeV}^4$\,; and 
$\sim42\,\%$\, for $V_1=0.5\,{\rm GeV}$\, and $G_2=0.0095\,{\rm GeV}^4$\,. 
 Concerning the masses in the above range of the anomalous mass defects, along the 
$V_1=0.5\,{\rm GeV}$ curve, they vary from $M\simeq 4.7\,M_\odot\,$ (maximum at the 
$\Delta_2M=0$ limit) at $G_2=2.5\times10^{-5}\,{\rm GeV}^4$ to $M\simeq0.63\,M_\odot\,$ at 
$G_2=0.0095\,{\rm GeV}^4$. 
For other values of $V_1$ the masses assume intermediate values.

 As $G_2$ increases, the apparent "convergence" of the curves led us (speculatively) to 
extrapolate our calculations to $G_2=0.1\,{\rm GeV}^4$, beyond the limits of the analysis 
made in Ref.\,\cite{Iof}, as shown in panel (b). 
 As a result we obtained a slightly ascending tail with 
$\Delta_2M\simeq-2.7\times10^{53}\,{\rm erg}$ at the endpoint. 
 Along this tail the masses vary, for example, 
from $M\simeq0.49\,M_\odot\,$ at $V_1=0$ and $G_2=0.05\,{\rm GeV}^4$ 
to $M\simeq0.25\,M_\odot\,$ at $V_1=0.5\,{\rm GeV}$ and $G_2=0.1\,{\rm GeV}^4$. 
 For the intermediate values of $V_1$ the masses are also intermediate.
 The mass excess fractions may be as large as 
$\sim33\,\%$\, for $V_1=0$ and $G_2=0.05\,{\rm GeV}^4$\,; 
$\sim52\,\%$\, for $V_1=0.3\,{\rm GeV}$ and $G_2=0.08\,{\rm GeV}^4$\,; and 
$\sim60\,\%$\, for $V_1=0.5\,{\rm GeV}$ and $G_2=0.1\,{\rm GeV}^4$\,.
 Extending (now, arbitrarily speculatively) our extrapolation to $G_2=1\,{\rm GeV}^4$\,, the 
anomalous mass defects along the tail are not greater than $-1.2\times10^{53}\,{\rm erg}$\,. 
 At $G_2=1\,{\rm GeV}^4\,$, the masses vary from $M\simeq0.11\,M_\odot\,$ at $V_1=0$ to    
$M\simeq0.09\,M_\odot\,$ at $V_1=0.5\,{\rm GeV}\,$; the excess fractions changing from 
$\sim68\%$ to $\sim74\%$\,, respectively.

 For large values of $G_2$ (say, $G_2>0.07\,{\rm GeV}^4$)\,, the curves concentrate in a 
narrow band.
 Then, it should be very difficult (or a very accurate determination, as for instance 
in gamma-ray bursts observations, should be required) to extract, from the $|\Delta_2M|$ 
measurements around  $\sim3\times10^{53}\,{\rm erg}$\,, reasonable estimates for $V_1$ and/or 
$G_2$. 
 By the way, as a general case in the $G_2-\Delta_2M$ plane, we need an additional 
measurement to determine unambiguously the model parameters $G_2$ and $V_1$.
 For the sake of comparison with the BM, the open circles along the $V_1=0$ curve correspond 
to the values of $B$ used in Refs.\;\cite{AFO,HZS,Koh,Var1,Var2,Var3,Ber}. 

 Here, a curious fact comes from the supernova SN1987A. 
 The neutrino signals were detected at Kamiokande II\,\cite{Kah} and at IMB\,\cite{Arn}. 
 The total energy of the observed neutrinos was found to be $\sim3\times10^{53}\,{\rm erg}$. 
 It was pointed out that one signal might have been originated in the formation of a 
neutron star after the supernova explosion and the other signal in a possible formation of a 
strange star \cite{Var3}.  
 It is inquisitive that the values of $|\Delta_2M|$ along the tail in panel (b) of 
Fig.\,\ref{d2mg2v11} are roughly coincident with the energy of the SN1987A neutrinos.
  Also, of similar magnitudes are the gamma-ray bursts GRB970828 with 
$\sim2.7\times10^{53}\,{\rm erg}$\,\cite{Ber} and  GRB971214 with an inferred energy loss of 
$\sim3\times10^{53}\,{\rm erg}$\,\cite{Kul} (assuming isotropic emissions).  
 Hence, measurements of $|\Delta_2M|$ around these energy values may be of particular importance.
 
 Finally, within our freedom to choose the value of $m_{\rm A}$, we performed our investigation  
assuming $m_{\rm A}=m_{\rm n}$.  
 Taking into account that $m_A$ enters the expression of $M_A$ as an external factor multiplying 
$N_A$ in Eq.\,\ref{NA}, the conversion formula given $\Delta_2M$ in terms of 
$m_0=m({\rm ^{56} F_{\rm e}})/56$ is 
\bqn 
\Delta_2M({\rm ^{56} F_{\rm e}})&=&\Delta_2M+(\eta-1)M_A\nonumber\\
&=&\Delta_2M-0.175\times10^{53}(M_A/M_\odot)\,,
\lb{d2MFe}
\eqn
where $\eta=m_0/m_n\simeq0.9902$ and 
$M_\odot\simeq1.988\times10^{33}{\rm g}\simeq1.787\times10^{54}{\rm erg}$\,. 
 In Fig.\,\ref{d2mg2v1Fe}, the curves corresponding to $m_A=m_0$ are slightly shifted with 
respect the ones for $m_A=m_{\rm n}$. 
 In the context of baryonic stars, interesting comments about the possibility of states 
with $\Delta_2M<0$ be reached by the release of nuclear energy, for the case of rarefied 
hydrogen and the case of rarefied iron vapor dispersed at infinity, are made in Ref.:\,\cite{ZeN}.

\section{Final remarks}
\lb{frmks}

 Previous investigations showed that the FCM provides new possibilities for the investigation 
of compact stellar configurations\,\cite{Bal,Bur,Plu,Bom,Log1,Log2,Fla1}.
 In the present work, we considered the general aspects of the mass defects of strange stars 
within the FCM without magnetic field, with special emphasis on the anomalous mass defects. 
 Concerning the magnitudes of the anomalous mass defects, our results are consistent with the 
estimated electromagnetic energies of the gamma-ray bursts, varying from $0.07\times10^{53}\,{\rm erg}$\, 
in  GRB970508 to $5\times10^{53}\,{\rm erg}$\, in  GRB011211 (assuming isotropic emission), 
given in Table 1 of Ref.\,\cite{Ber}\,, and the theoretical energy predictions of quark-nova 
explosions \cite{RaD}.

 Since quarks are not freely observed, a strange star in an explosion process must liberate 
its energy excess as neutrinos, gamma-rays, gravitational waves, or other forms of matter/energy.   
 If the ejected matter is made of hadrons, then a transformation to the hadron phase is needed 
in order to disperse the hadronic content to infinity. 
 Another possibility would be the energy release in the form of strangelets \cite{Wit,Far,AFO}. 
 Strangelets are lumps of self-bound matter containing as few as a thousand of u, d and s quarks. 
 The question of strangelets was considered in Ref.\,\cite{Klu}. 
 It was argued that a disruption of a strange star would contaminate the Galaxy with an exceeding 
density of strangelets with respect to that required to transform neutron stars into strange 
stars\,\cite{Cal}.
 However, in order to such a contamination takes place, it is expected that the strangelets 
must survive a long time, hence the need to investigate the SQM stability.
  
 Very recently, we considered for several values of $V_1$ the behavior of the stability 
windows of the SQM (with respect to the $^{56}F_\ex$ nucleus) with chemical equilibrium 
and charge neutrality (cf. panel (a) of Fig.\,3 in Ref.:\,\cite{Fla2}). 
 Within the same line, we determined here the value of $V_1$ in order to give a zero stability 
window at the s-quark mass $m_s=0.15\,{\rm GeV}$\,.
 As a result we obtained $V_1=0.33\,{\rm GeV}$, as shown in panel (a) of Fig.\,\ref{fewin}, 
which also includes (for comparison) the nonzero windows at $V_1=0$\,, 0.1 GeV and 0.2 GeV. 
 At the given s-quark mass (dashed horizontal line), the stability windows are zero in the range 
$0.33\,{\rm GeV}\leq V_1\leq0.5\,{\rm GeV}$\,. 
 On the other hand, the largest window width occurs at $V_1=0$ for any s-quark mass $m_s\geq0$, 
being maximum at $m_s=0$\,. Then, it should be interesting to express the SQM stability, at a 
given $m_s$, in terms of a relative stability, which we crudely estimated by considering 
three different possibilities.
 If for each value of $V_1$\,, we compare the width of the stability window at 
$m_s=0.15\,{\rm GeV}$ with the corresponding (maximum) value at $m_s=0$\,, we observe 
that the SQM stability gradually decreases from about 53\,\% at $V_1=0$ to zero at 
$V_1=0.33\,{\rm GeV}$, as shown in panel (b).
 On the other hand, if for each value of $V_1$\,, we compare the width of the stability 
window at $m_s=0.15\,{\rm GeV}$ with the one at $V_1=0$ (at the same $m_s=0.15\,{\rm GeV}$), 
the SQM stability also decreases very rapidly from 100\,\% at $V_1=0$ to zero at 
$V_1=0.33\,{\rm GeV}$\,(short dashed line).
 Finally, if for each value of $V_1$\,, we compare the width of the stability window at 
$m_s=0.15\,{\rm GeV}$ with the largest one (at $m_s=0$ and $V_1=0$)\, we observe the 
fastest rate of stability decrease (long dashed line).
 The relative  stabilities are $\gappl40\%$ at $V_1\simeq0.1\,{\rm GeV}$\,; $\gappl20.5\%$ 
at $V_1\simeq0.2\,{\rm GeV}$\, and $\gappl2.5\%$\, at $V_1\simeq0.3\,{\rm GeV}$\,. 
 Then, in this aspect, for reasonable values $V_1$ it is unlikely that the strangelets 
ejected from strange star explosions should survive so long (before they decay) to arrive 
on Earth or other place of the Galaxy.
 These results appeared to be in accordance with terrestrial experiments at RHIC which 
have not confirmed the existence of the SQM nor proved that it does not exists
\,\cite{San,Abe,Ble}. 

 In an investigation relating the gamma-ray bursts to a second explosion after the (first) 
supernova explosion, it was pointed out that the main difficulties of the model were to 
explain the causes of the second explosion and the time elapsed between the first explosion  
and second explosion\,\cite{Bom2}.
 In this regard, another interesting aspect to be considered would be the connection between 
the anomalous mass defects and stellar instabilities. 
 Merely speculatively, it appears to be reasonable to expect that the greater the magnitude of 
the anomalous mass defect is, the greater might be the instability of a strange star configuration. 
 Correspondingly, the lower might be the time interval between the first (supernova) explosion 
and the second (quark-nova) explosion. 
 However, such an investigation should require detailed studies of the internal structure as 
well as internal processes of the strange stars, that merit to be considered elsewhere, in the 
author's opinion.  

 Finally, although the main aim of the present work is the study of the mass defects of the 
strange stars, with emphasis on the anomalous mass defects, let us here consider some general 
aspects concerning the low-mass strange stars as well as their importance for further 
investigations in the FCM.

\begin{figure*}[th]
\centerline{
\psfig{figure=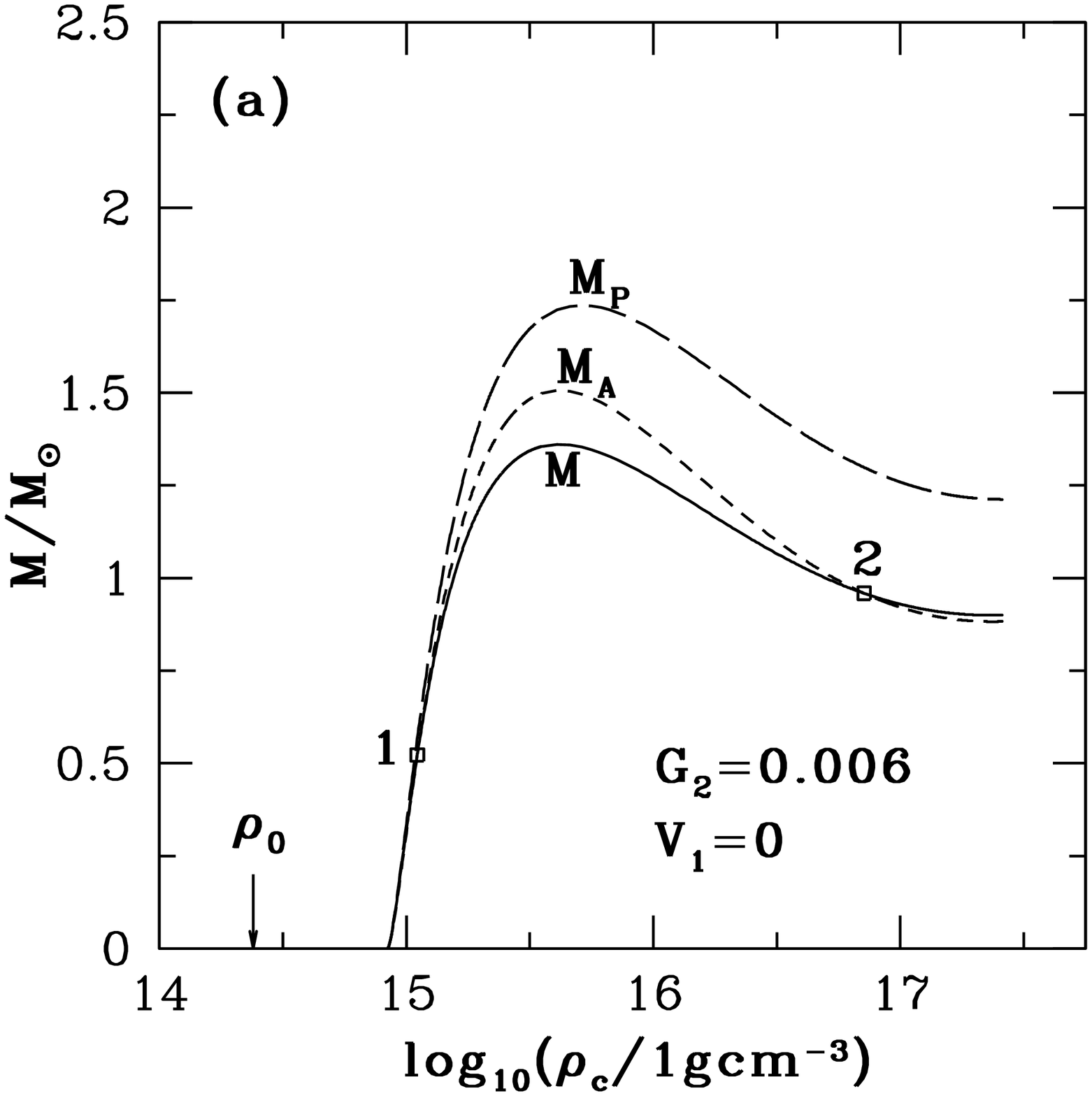,width=3.2truein,height=3.2truein}
\psfig{figure=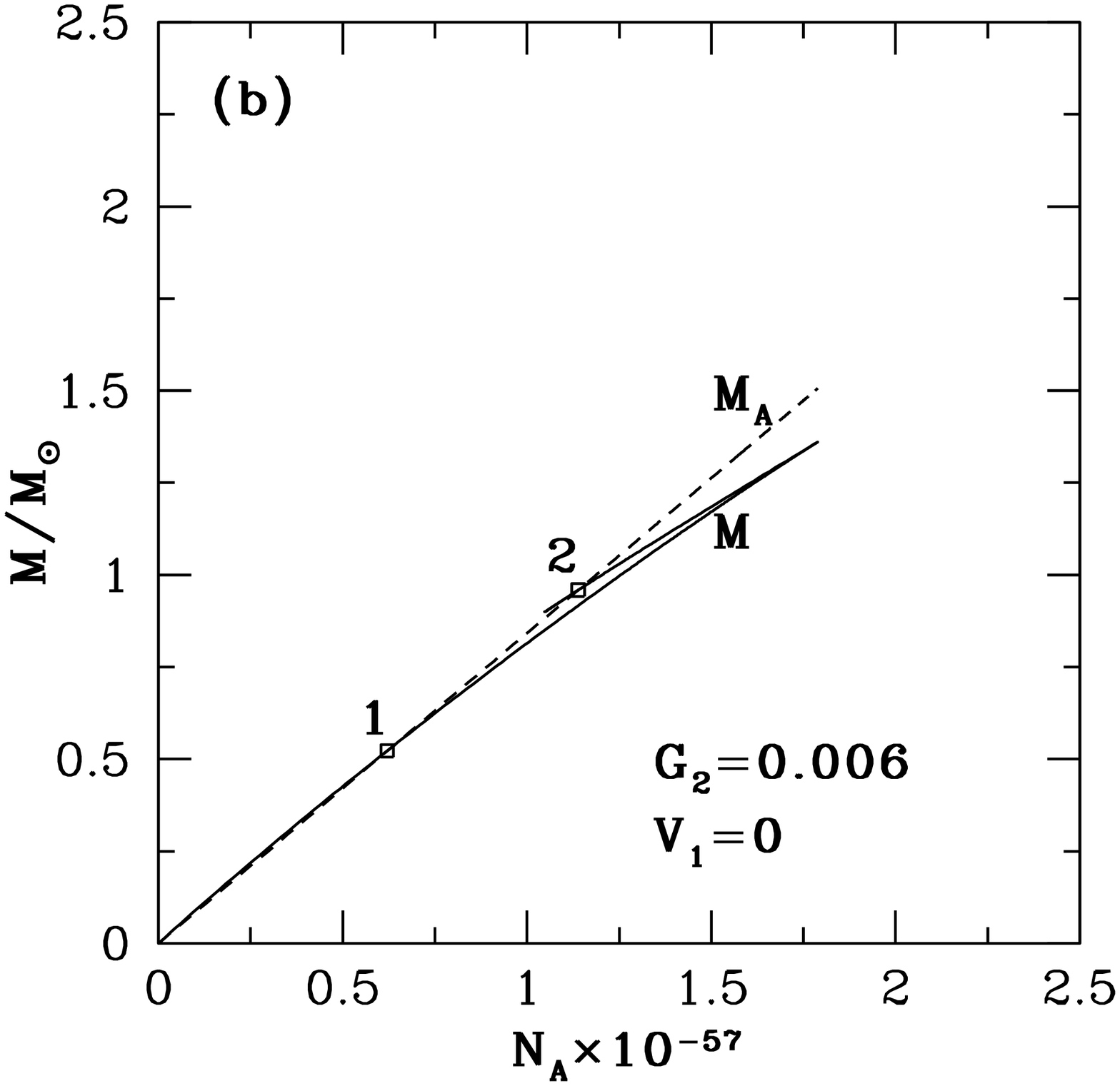,width=3.2truein,height=3.2truein}
\hskip .5in}
\centerline{
\psfig{figure=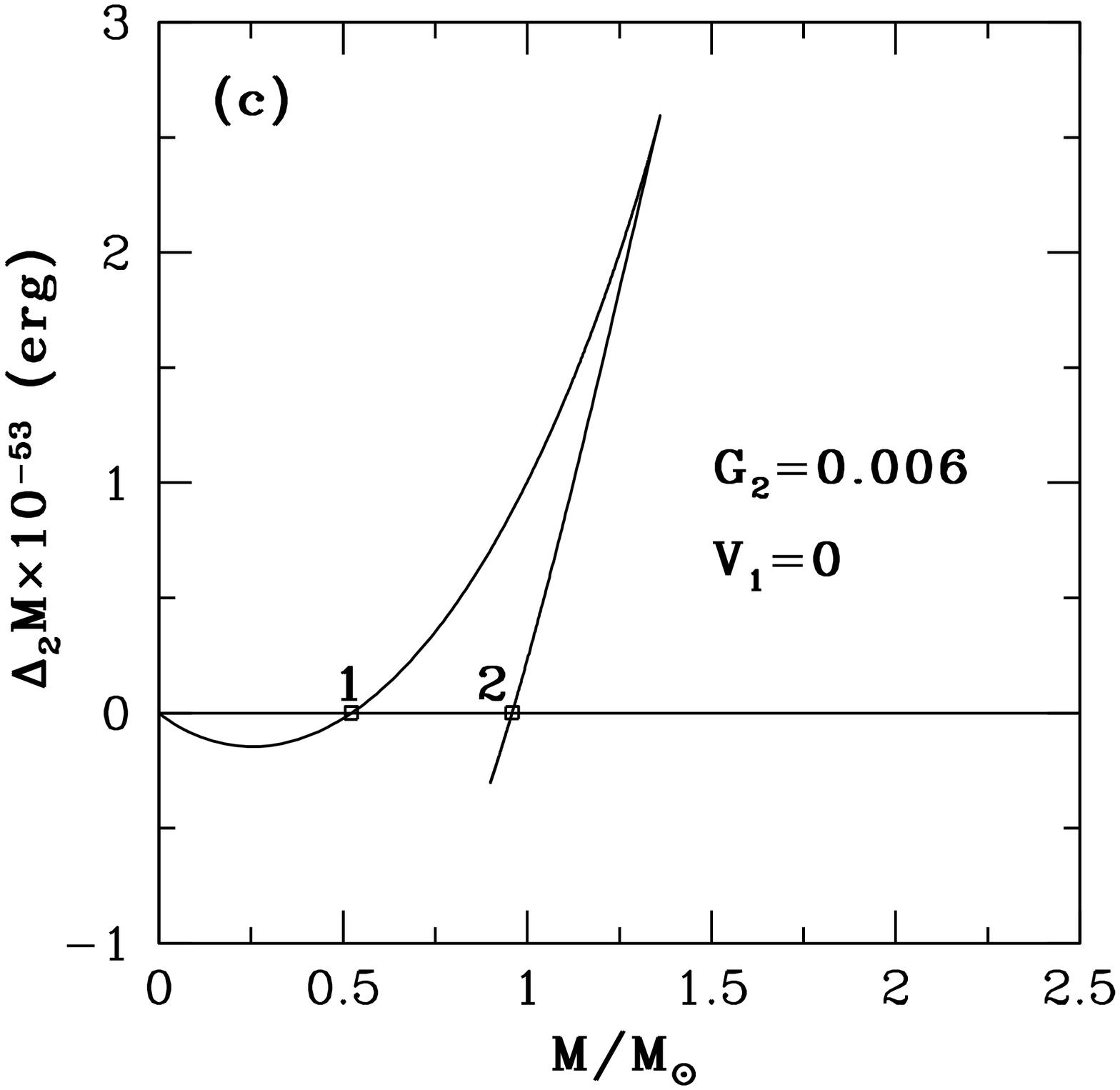,width=3.2truein,height=3.2truein}
\hskip .5in}
\caption{ 
}
\label{mg2v11}
\end{figure*}

 In Sec.\,\ref{res}, we showed that low-mass strange stars with anomalous mass defects are 
possible in the first branch, as in Fig.\,\ref{mg2v11}.
 These stars are important for both theoretical and observational investigations. 
 For $M>M_\odot$, strange stars and neutron stars with the same mass present similar radii, 
but for $M<M_\odot$ their radii are markedly different\,\cite{AFO,XDL1}. 
 Another feature is that the strange stars, being more compact, have lager surface redshifts 
than the ones for neutron stars\,\cite{NS1}.
 It would be possible to distinguish strange stars and neutron star by radii direct measurements 
of low mass pulsar-like stars by observations from X-ray satellites\,\cite{Xu1}(and references 
therein).
 The identification of strange stars with $M\gappl0.1M_\odot$ can show us if they are bare 
stars given that their radii are much lower than the ones for stars with crust, in the low mass 
limit\,\cite{Xu2}.

Due to the fact that the quarks and anti-quarks are held together by the strong interaction forces, 
the strange stars are self-bound systems, even  in the absence of gravitation, whereas neutron 
stars are not. 
 In the low-mass regime, the mass-radius relations of strange stars have been well represented 
by the use of the approximated BM equation of state in the $m_q\rightarrow0$ (q=u,d,s) limit, 
\bq
p=\frac{1}{3}(\varepsilon-4\,B)
\lb{lmBMp}
\eq
which was used to calculate mass-radius relations in Refs.\,\cite{Wit,AFO}.

 The low-mass strange stars obey the $M\propto R^3$ dependence which can be easily explained 
within the BM.
 For $M\gappl0.3M_\odot$\,, the gravitational pull becomes small compared to the contribution of 
the vacuum energy density represented by the bag constant $B$. 
 Inside the star, due to the high degree of incompressibility of the SQM, the energy density is 
nearly constant (cf. Fig.\,1 in Ref.\,\cite{AFO} or Fig.\,8.6 in Ref.\,\cite{NS1}. 
 In this case, the Newtonian approximation suffices to calculate the mass of the strange star 
as being $M\simeq(4\pi/3)R^3\,\varepsilon$\,.

 In the FCM, the corresponding  simplified NPEOS is (from Eqs.\,(\ref{pqT0})-(\ref{dvac})) given by 
\bqn
p&=&\frac{1}{3}\bigg[\varepsilon - \frac{3}{2}V_1n_A - 4\Delta|\varepsilon_{\rm vac}| \bigg] \nonumber\\
&=&\frac{1}{3}\bigg[\varepsilon - \frac{3}{2}V_1n_A - \frac{9}{16}G_2 \bigg]\,.
\lb{lowmp}
\eqn

\begin{figure*}[th]
\centerline{
\psfig{figure=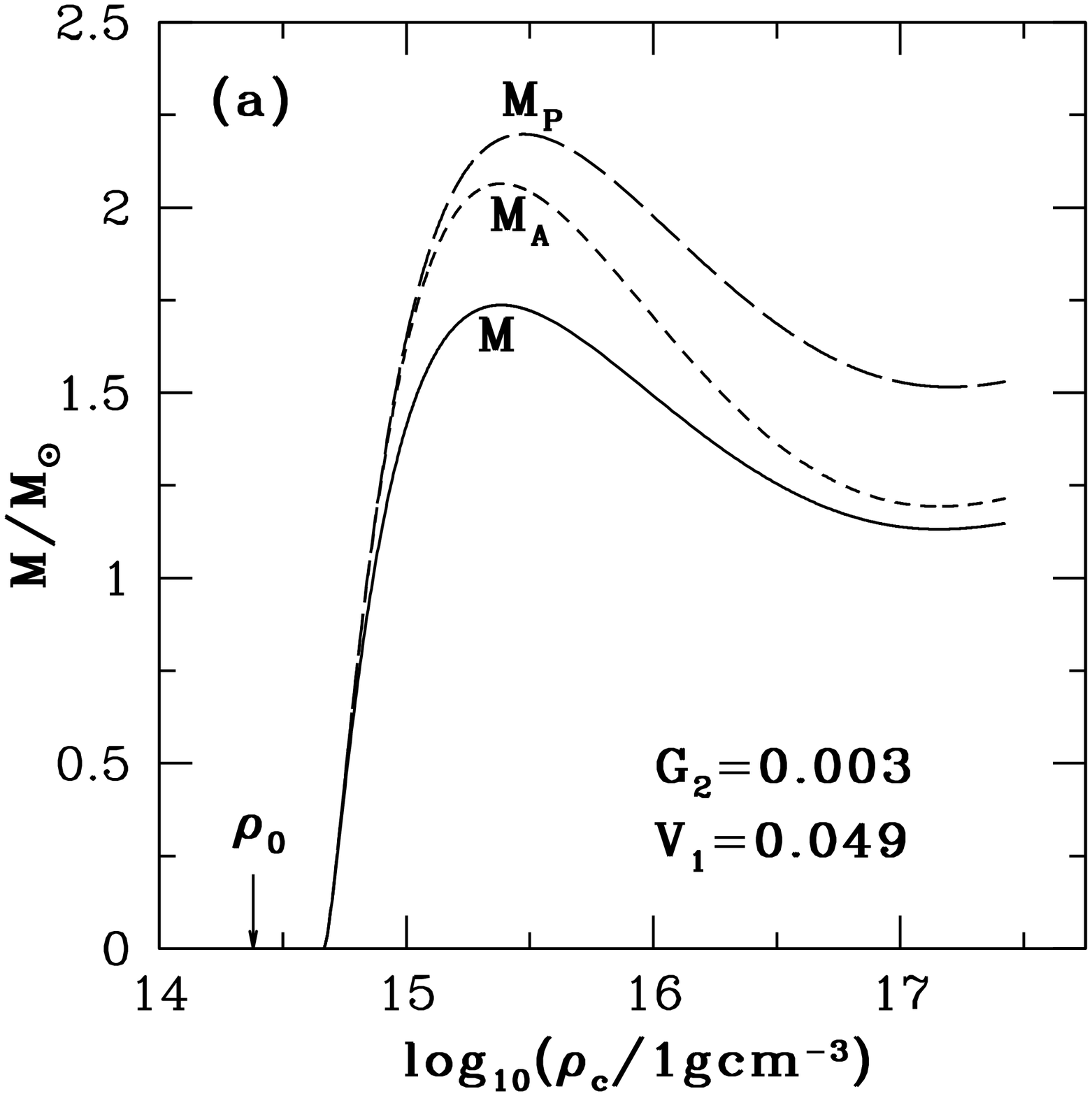,width=2.25truein,height=2.25truein}
\psfig{figure=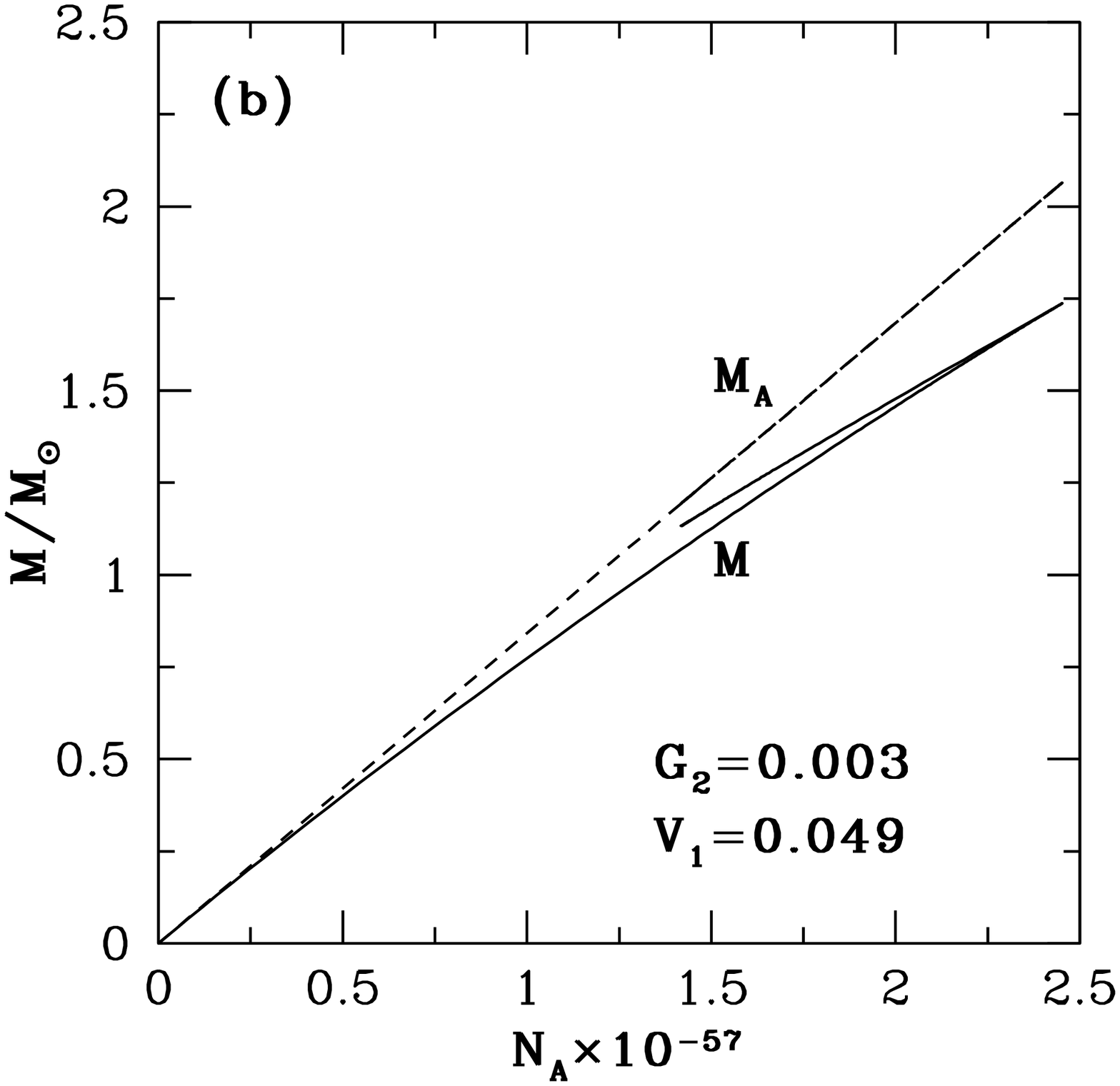,width=2.25truein,height=2.25truein}
\psfig{figure=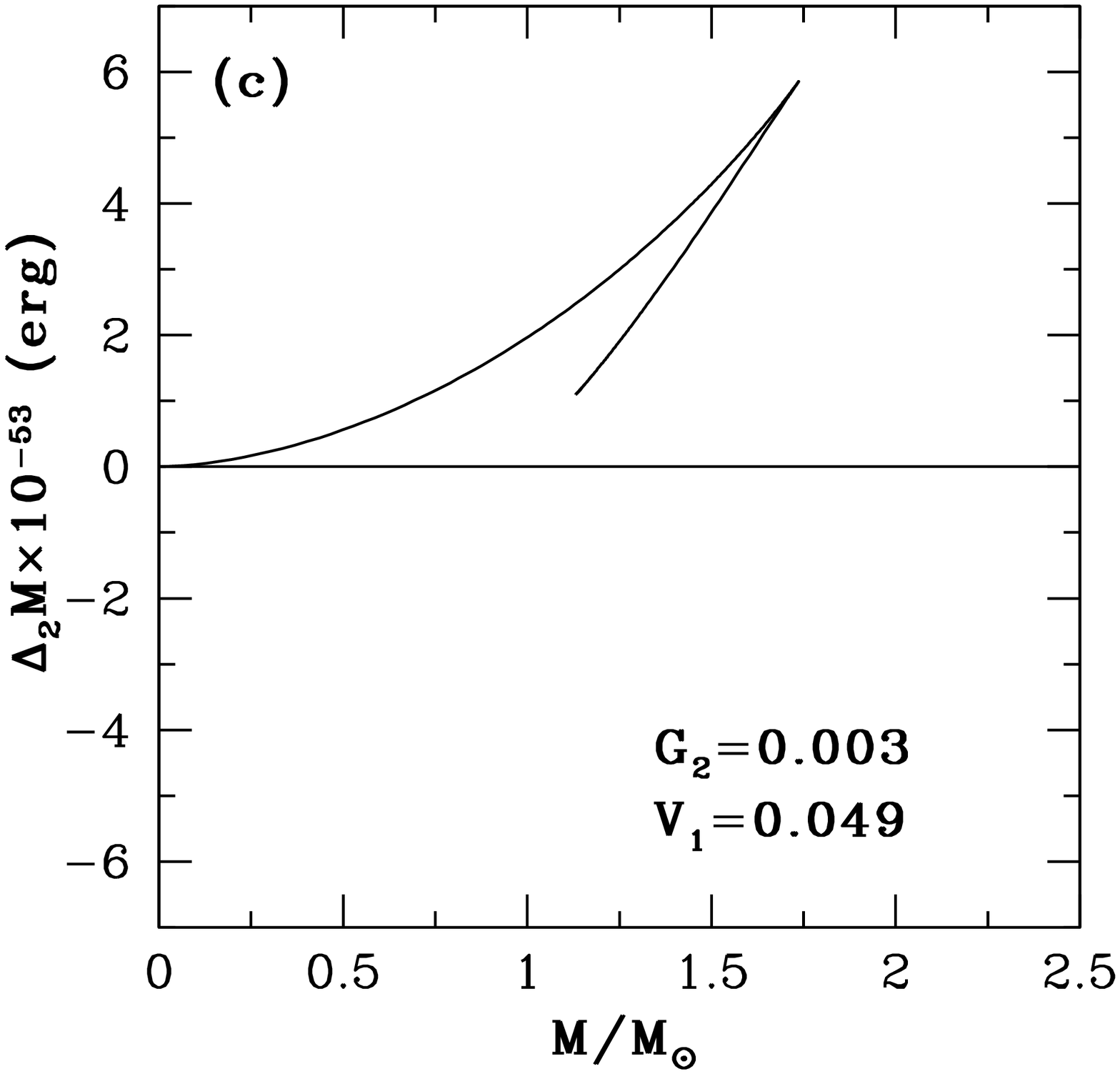,width=2.25truein,height=2.25truein}
\hskip .5in}
\centerline{
\psfig{figure=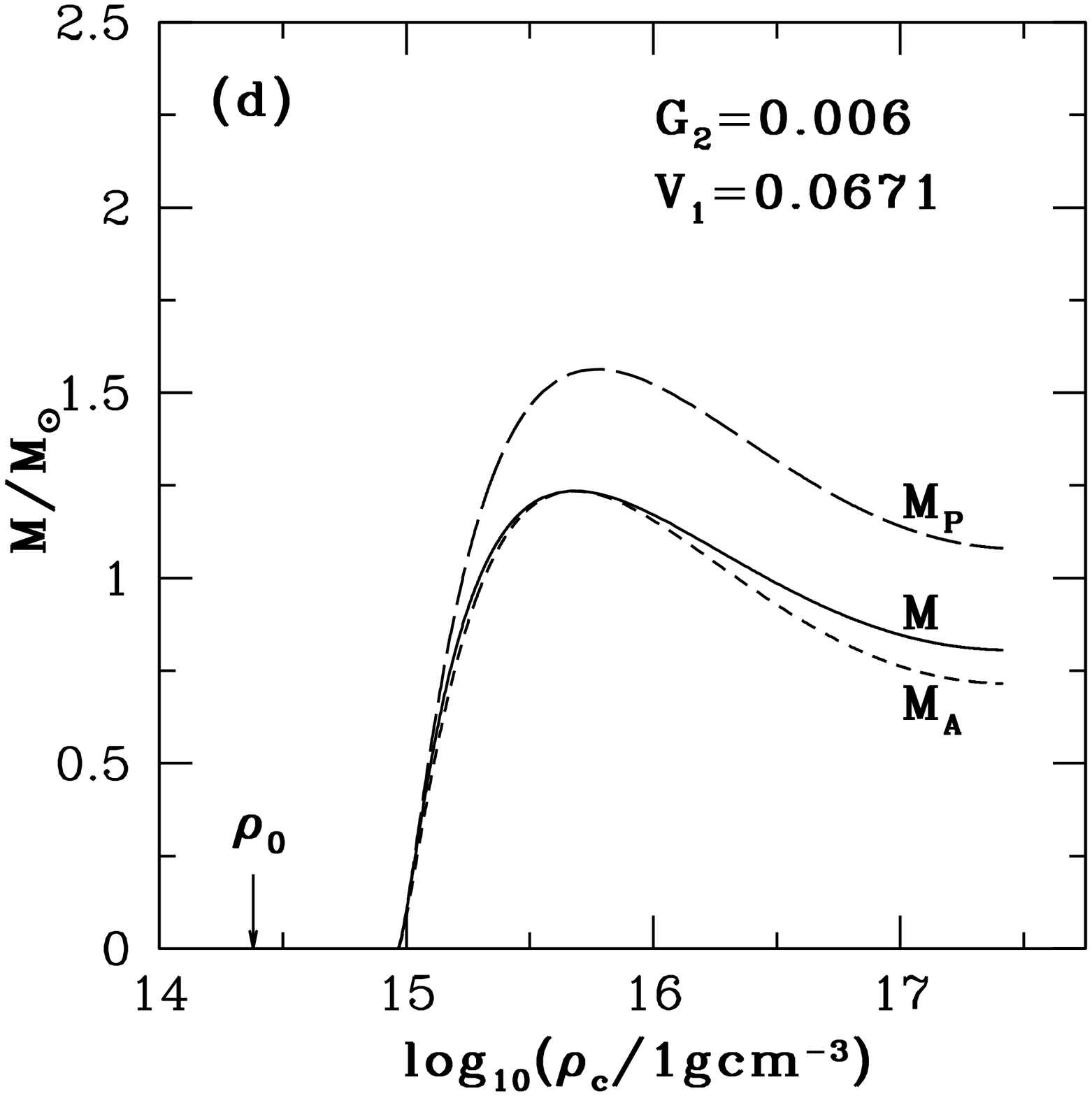,width=2.25truein,height=2.25truein}
\psfig{figure=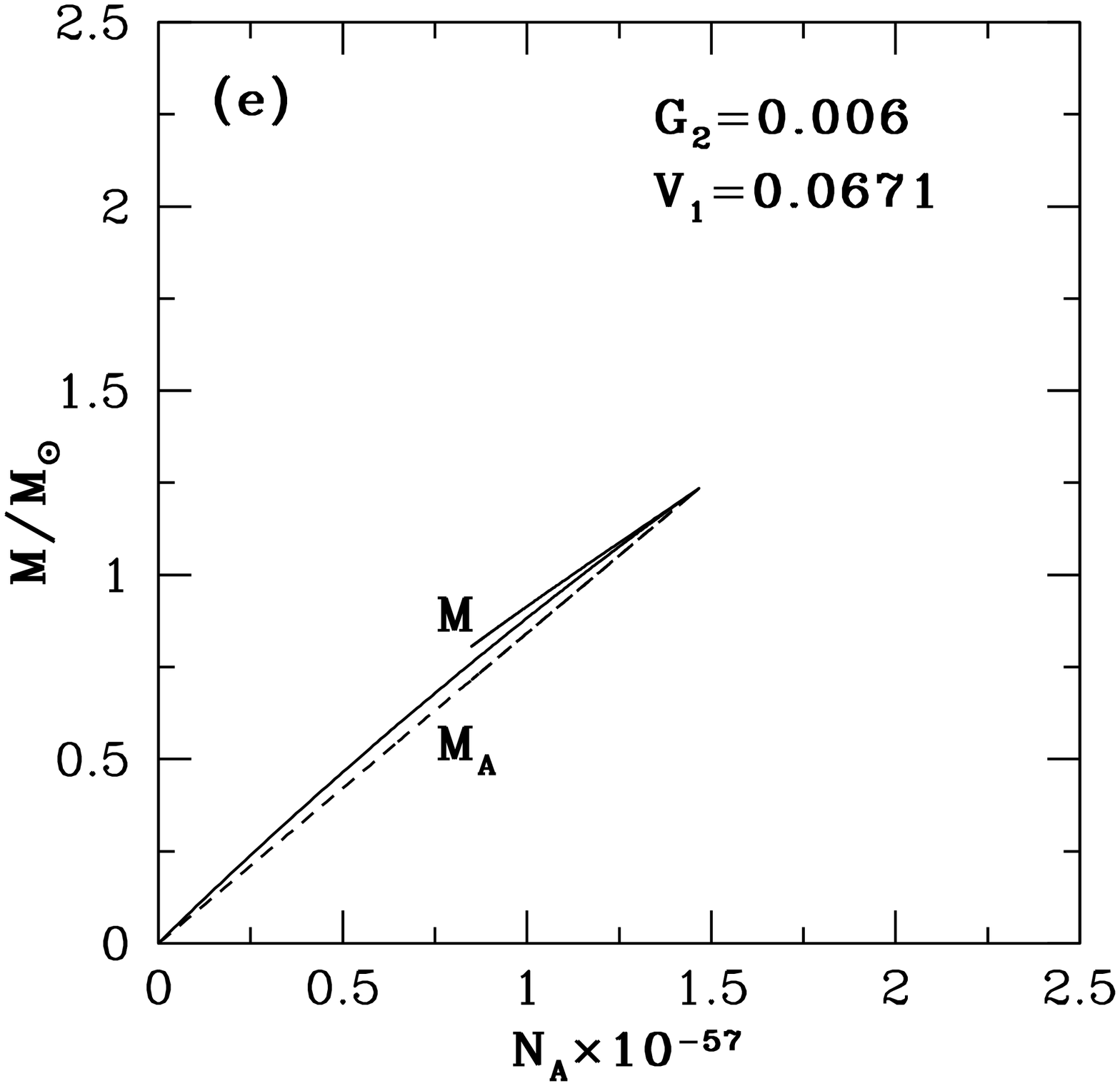,width=2.25truein,height=2.25truein}
\psfig{figure=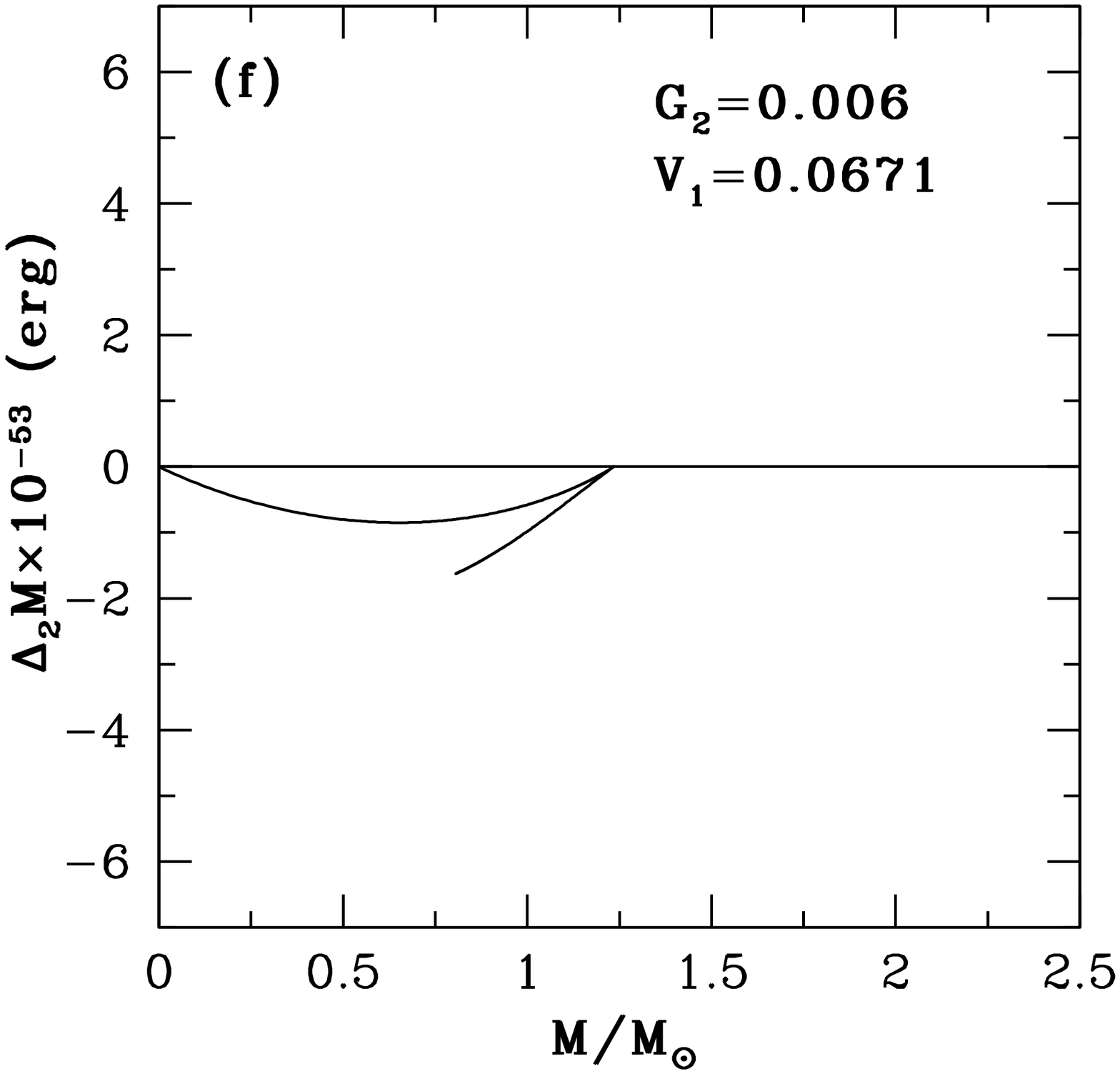,width=2.25truein,height=2.25truein}
\hskip .5in}
\centerline{
\psfig{figure=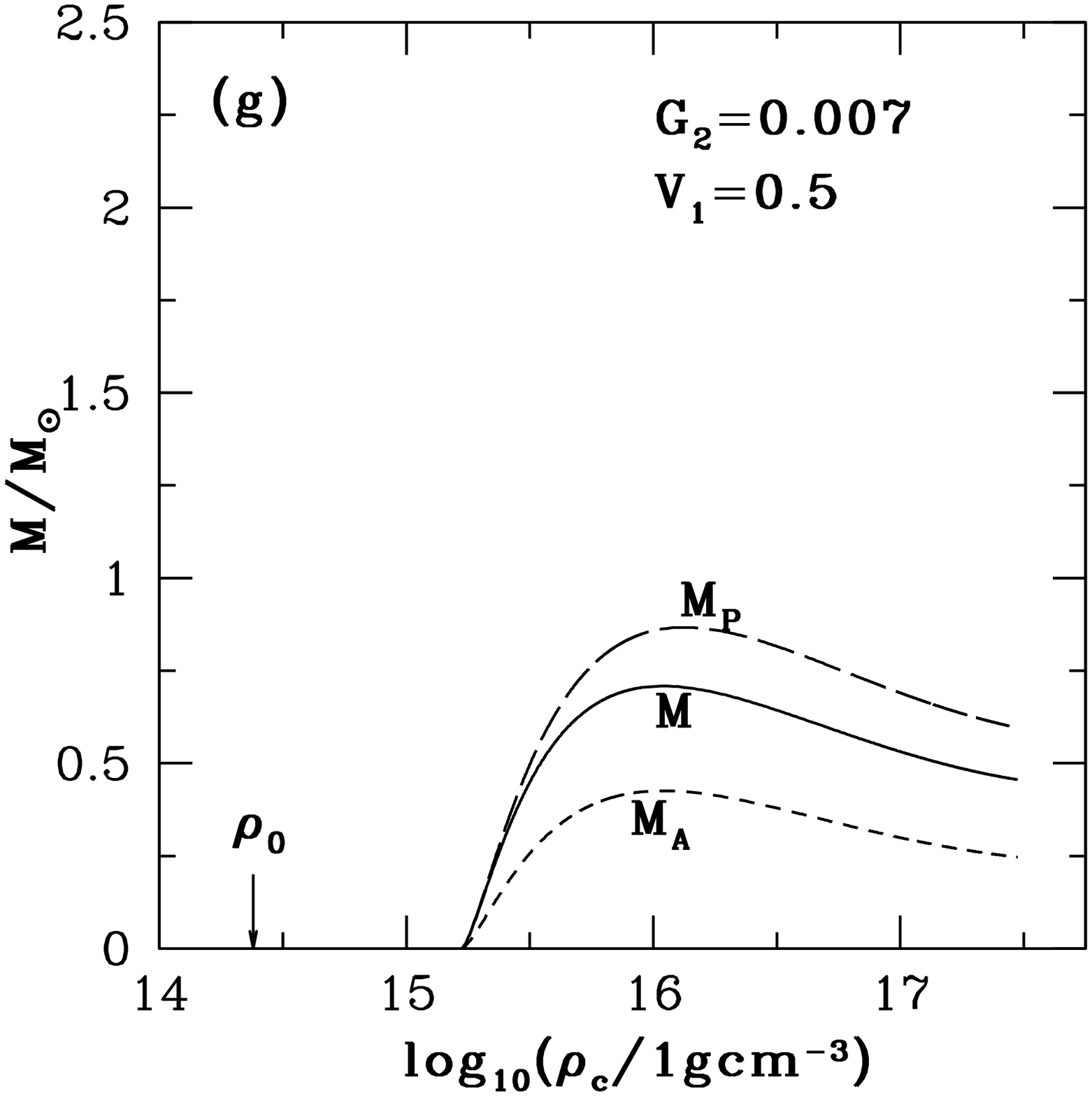,width=2.25truein,height=2.25truein}
\psfig{figure=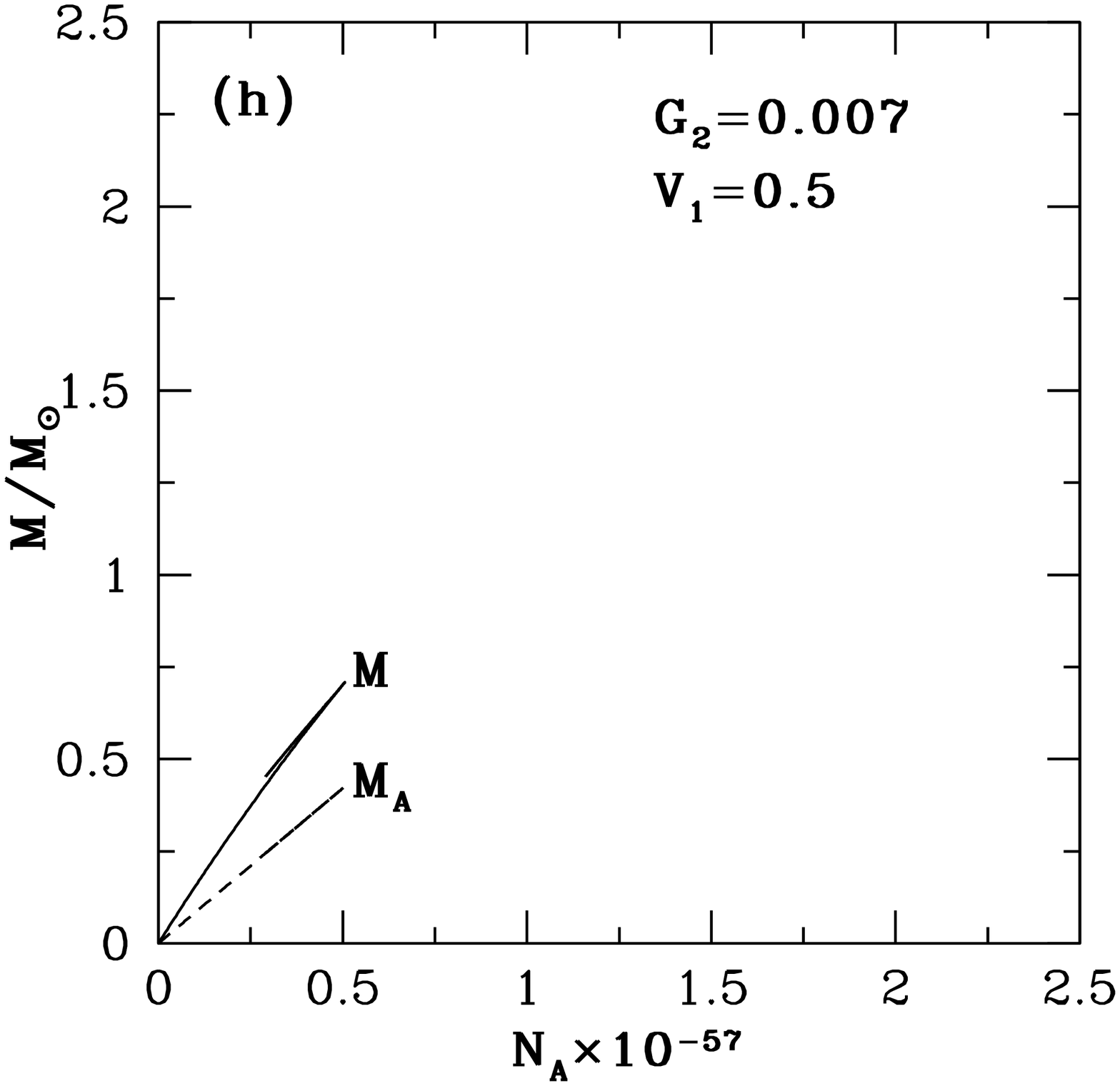,width=2.25truein,height=2.25truein}
\psfig{figure=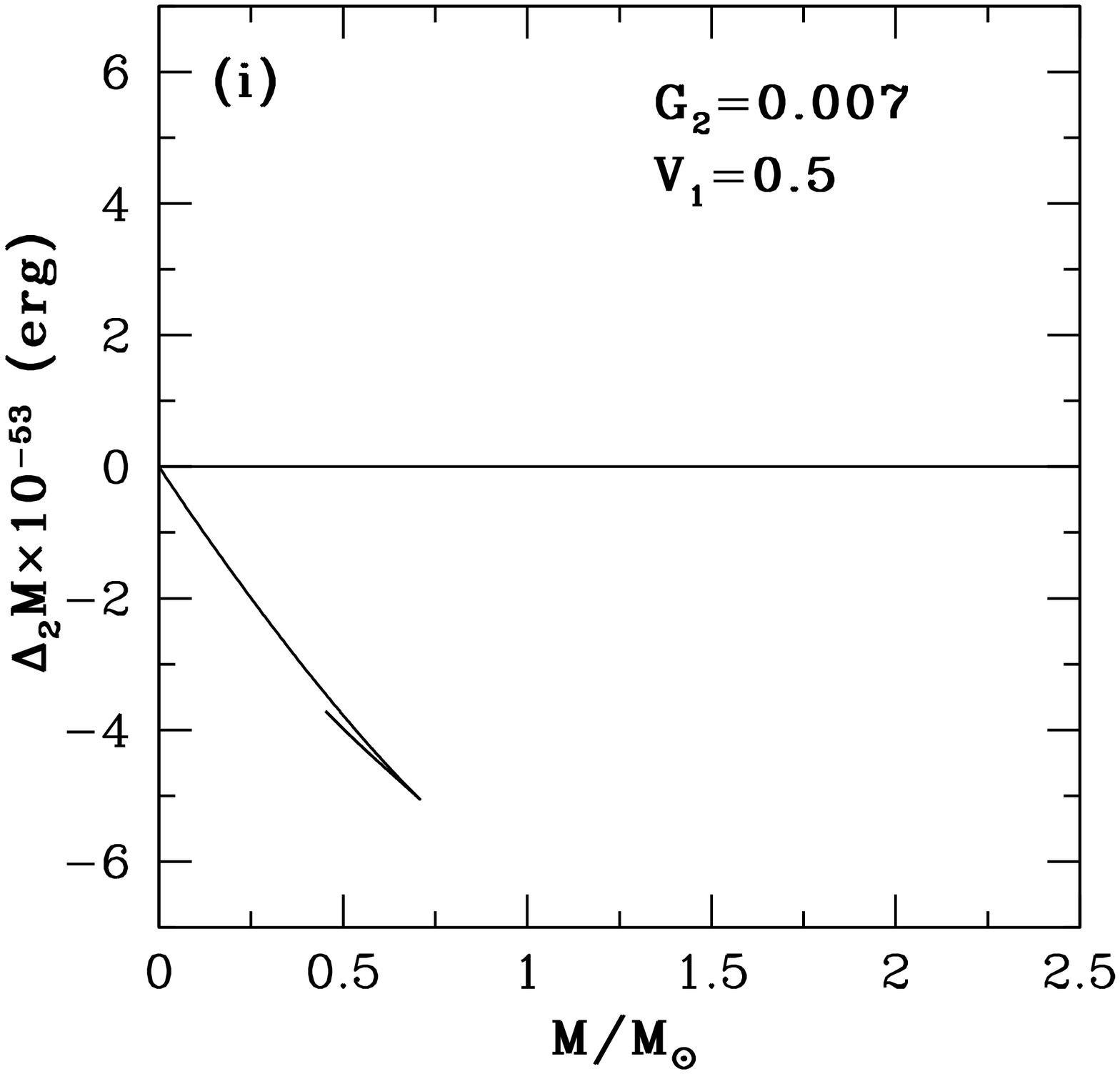,width=2.25truein,height=2.25truein}
\hskip .5in} 
\caption{ 
} 
\label{mg2v12}
\end{figure*}

\begin{figure*}[th]
\centerline{
\psfig{figure=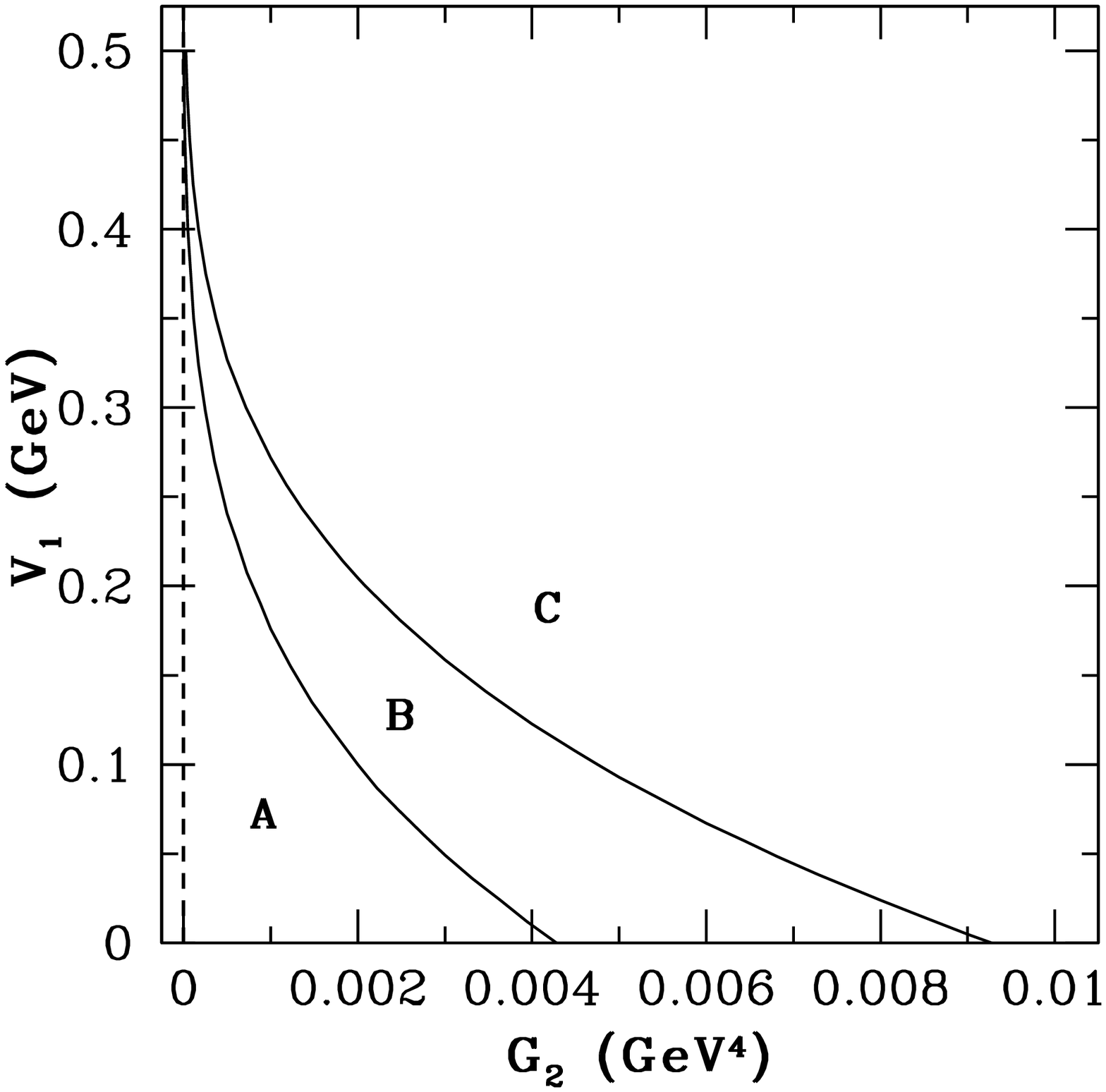,width=3.2truein,height=3.2truein}
\hskip .5in}
\caption{ 
}
\label{d2mg2v1}
\end{figure*}

\begin{figure*}[th]
\centerline{
\psfig{figure=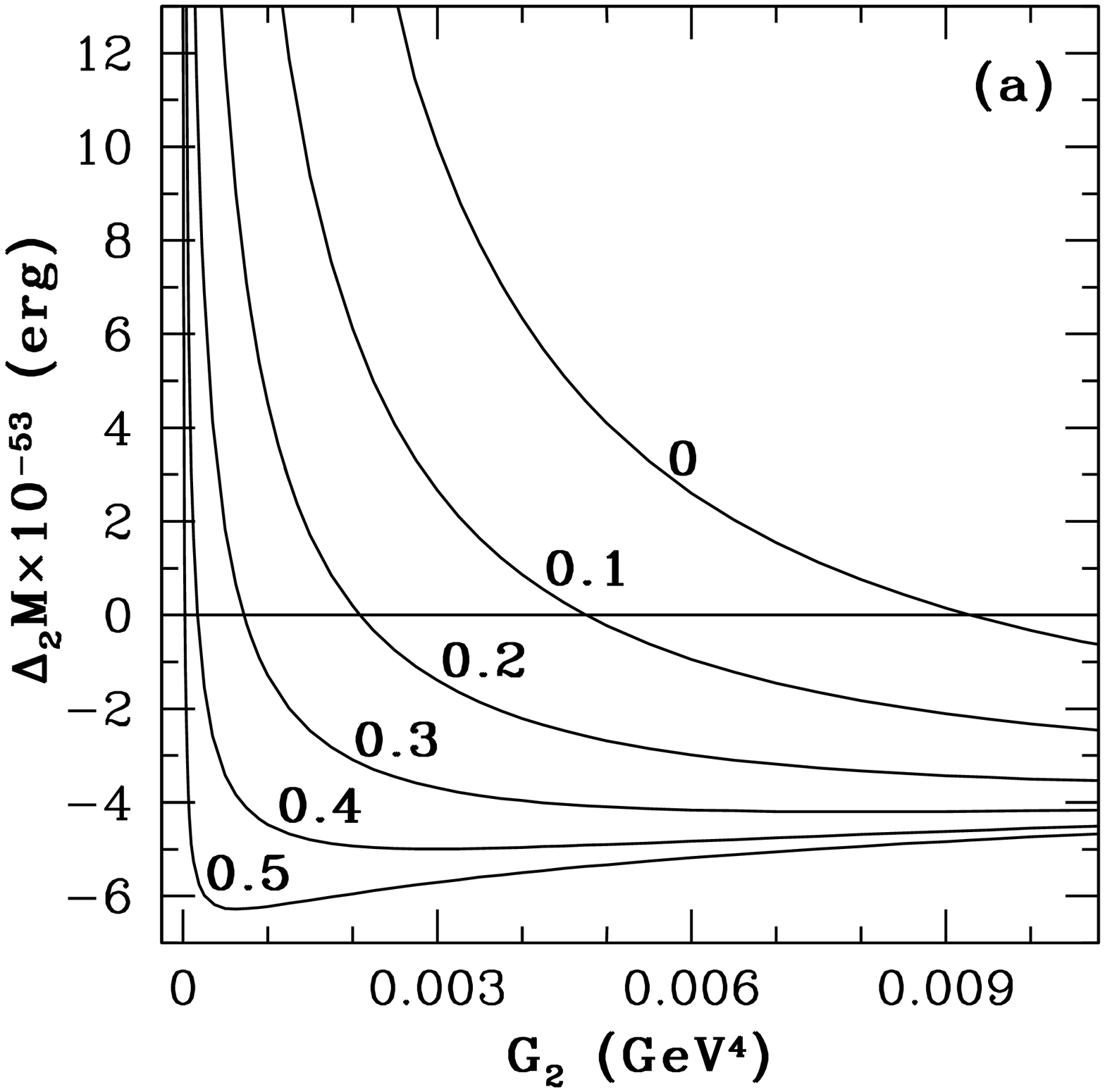,width=3.2truein,height=3.2truein}
 \psfig{figure=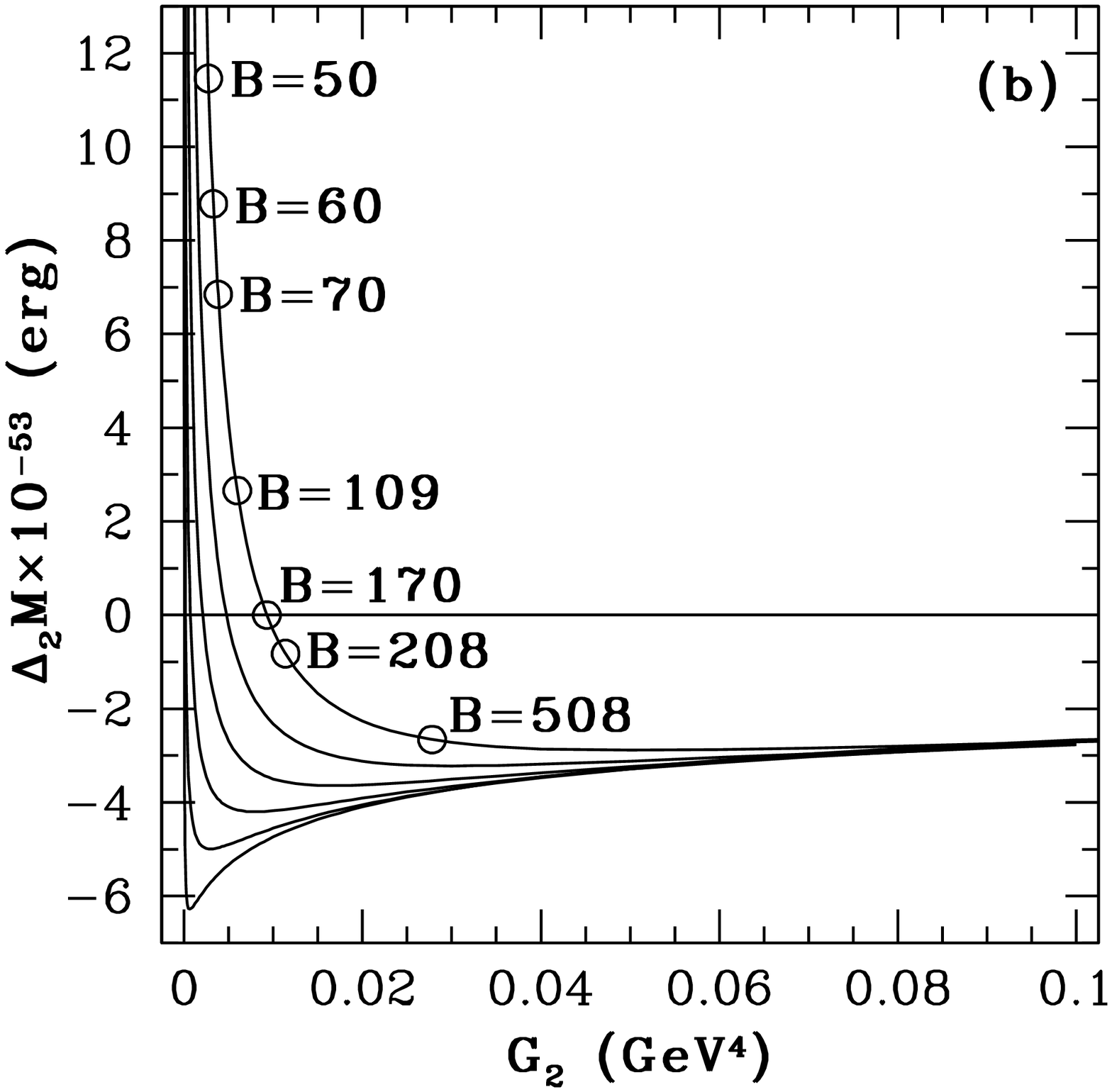,width=3.2truein,height=3.2truein}
\hskip .5in}
\caption{ 
}
\label{d2mg2v11}
\end{figure*}

\begin{figure*}[th]
\centerline{
\psfig{figure=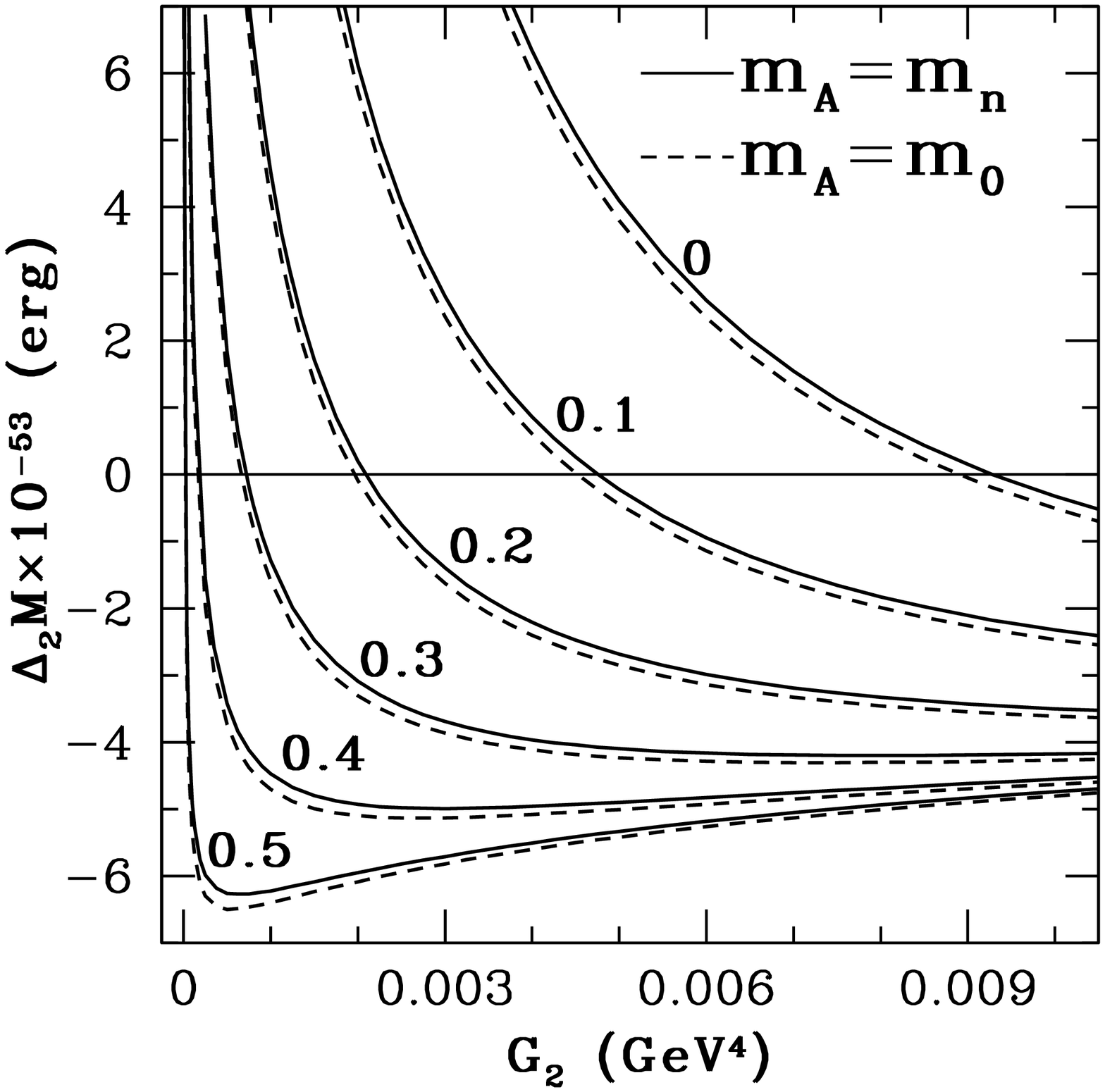,width=3.2truein,height=3.2truein}
\hskip .5in}
\caption{ 
}
\label{d2mg2v1Fe}
\end{figure*}

\begin{figure*}[th]
\centerline{
\psfig{figure=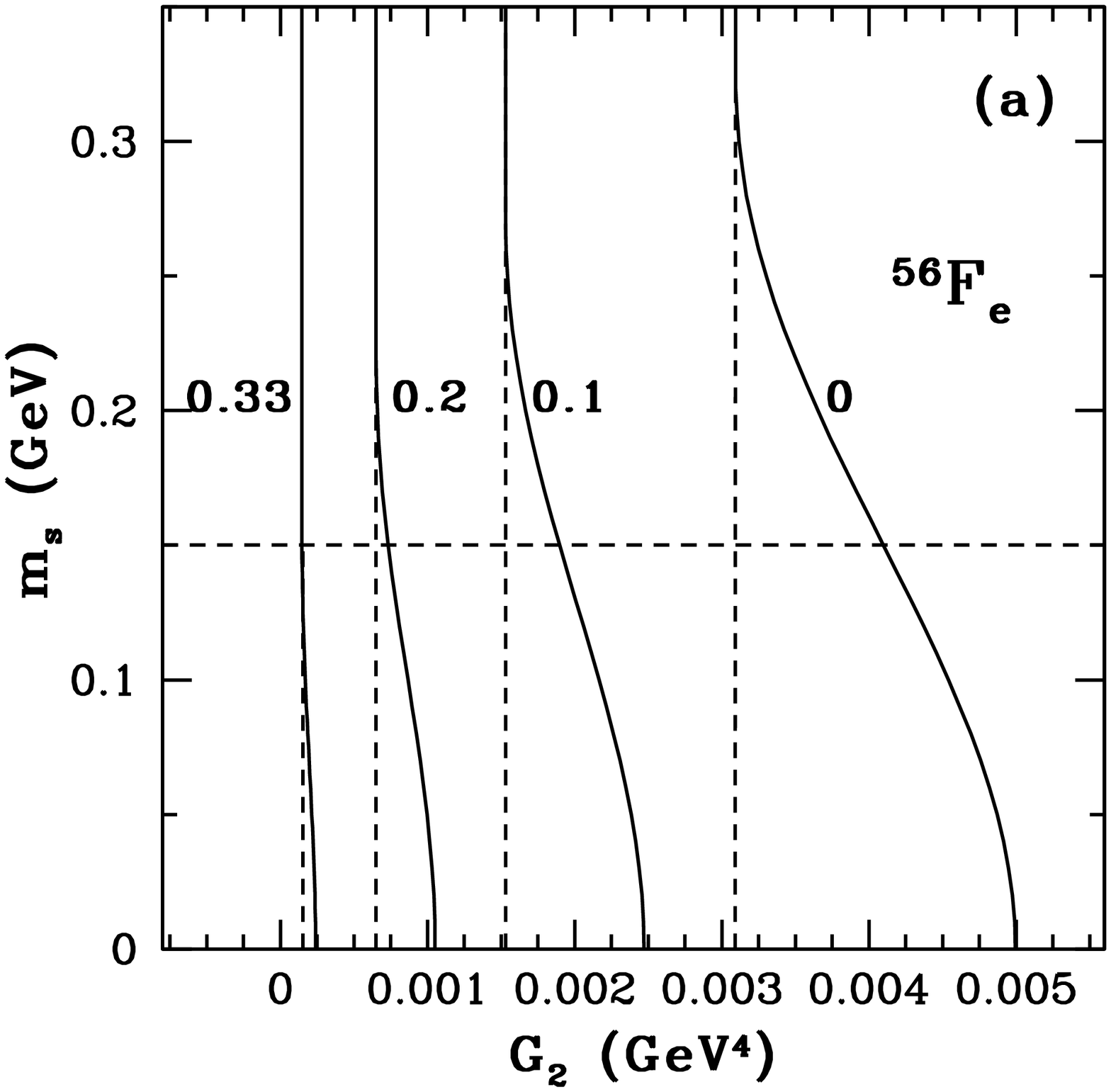,width=3.2truein,height=3.2truein}
\psfig{figure=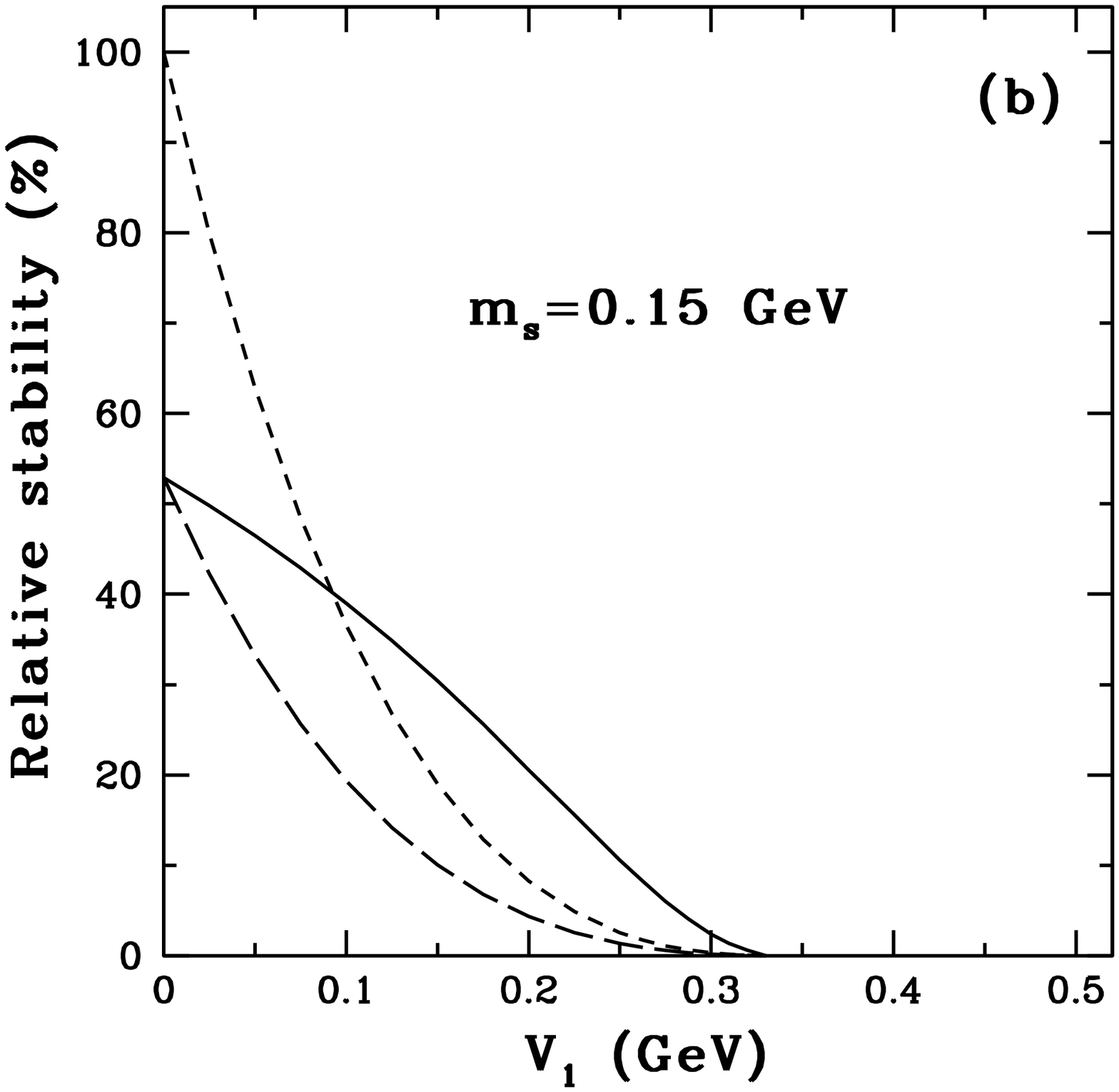,width=3.2truein,height=3.2truein}
\hskip .5in}
\caption{ 
}
\label{fewin}
\end{figure*}

 There is here an important point to be addressed. 
 Apart from the low-mass strange stars in the first branch, where the use of Eq.\,(\ref{lowmp}) 
is valid, we also have sequences of stellar configurations with maximum masses in the low-mass 
region (as in panel (g) of Fig.\,\ref{mg2v12}) where Eq.\,(\ref{lowmp}) is not generally valid, 
even when $M_{\rm max}\simeq0.3\,M_\odot$. 
 For instance, in the region of anomalous mass defects, corresponding to $0\leq V_1\leq0.5$\,GeV 
and $G_2$ in the extrapolated tail ($0.02\,{\rm GeV}^4\gappl G_2\gappl0.1\,{\rm GeV}^4$) in 
Fig.\,\ref{d2mg2v11}\,(panel (b)), the maximum masses are in the region 
$0.2\,M_\odot<M_{\rm max}<0.5\,M_\odot$\,.
 However, as a part of a next work, we say in advance that it is not generally true that 
Eq.\,(\ref{lowmp}) is an appropriate approximation\,\cite{Fla4}. 
 According to our preliminary exact numerical calculations, depending on the values of $V_1$ and 
$G_2$, inside a star with $M=M_{\rm max}$ in the low-mass region it is not mandatory for the energy 
density $\varepsilon(r)$ be nearly constant in order to allow for the usage of the Newtonian 
approximation.
 In this case, it appears that the concept of low-mass strange stars must be taken as meaning 
$M<M_{\rm max}$ rather than $M<M_\odot$\,. 
 Correspondingly, we expect that these features may imply interesting consequences to the anomalous 
mass defects.

\vspace{.2in}

\centerline{\bf ACKNOWLEDGMENTS} 

 I would like to thank J. S. Alcaniz, R. Silva and A. P. Santos.
 This work was done with the support provided by the Minist\'erio da Ci\^encia , Tecnologia 
e Inova\c c\~ao (MCTI).

\newpage

%
%

\newpage

\centerline{\bf FIGURE CAPTION}

{\bf Fig. 1 -}  For the given values of $G_2$ (in ${\rm GeV}^4$ units) and $V_1$ (in ${\rm GeV}$ units):  
 panel (a) - gravitational mass $M$ (solid line), baryonic mass $M_A$ (short dashed line) 
and proper mass $M_P$ (long dashed line) as functions of the central density. 
 Labels 1 and 2 indicate the points at which $M=M_A$, where the solid line and short dashed 
line cross itself.
 Panel (b): $M$ (solid line) and $M_A$ (dashed line) as functions of the baryonic number $N_A$\,.
 Panel (c): mass defect as function of $M$.
 In panels (b) and (c) the labels  1 and 2 correspond to the ones in panel (a). 
 All masses in units of the solar mass $M_\odot$\,.

{\bf Fig. 2 -}  As in Fig.\,\ref{mg2v11}, but for different values of $G_2$ and $V_1$ from top 
to bottom; each row with the same values of $G_2$ and $V_1$ from left to right. 

{\bf Fig. 3 -}  The $G_2$ - $V_1$ plane showing the different regions according to the sign of $\Delta_2M$.
 Region A: $\Delta_2M>0$ everywhere on the sequence of strange star configurations, as in 
panel (c) of Fig.\,\ref{mg2v12}.
 Region B: sequences with $\Delta_2M\geq0$ in the second branch and $\Delta_2M<0$ in the first 
branch or in both first branch and third branch, as in panel (c) of Fig.\,\ref{mg2v11}.
 Region C: $\Delta_2M<0$ everywhere on the sequence, as in panel (i) of Fig.\,\ref{mg2v12}. 
 The upper curve corresponds to $\Delta_2M=0$ at the maximum mass of the sequence, as in 
panel (f) of Fig.\,\ref{mg2v12} (cf.\;Fig.\;6 in Ref.\;\cite{Fla1}). 
 The lower curve  corresponds to the limit between the regions A and B, where $\Delta_2M=0$ 
at the first endpoint as in panel (c) of Fig.\,\ref{mg2v12}, or both first endpoint and second 
endpoint of the sequence.

 {\bf Fig. 4 -}  Panel (a): For several values of $V_1$, the mass defect at the maximum mass of 
the sequence of stellar configurations  as function of $G_2$. 
 Each curve is labeled by the value of $V_1$ (in ${\rm GeV}$ units) ranging from zero to 
$0.5$ GeV. 
 Panel (b): As in panel (a), but for values of $G_2$ extended up to $0.1\,{\rm GeV }^4$. 
 The open circles at the upper part of the $V_1=0$ curve correspond to some values of 
$B\,({\rm in\,MeV\,fm^{-3}})$ extracted from  the literature between $B=50\;{\rm MeV\,fm^{-3}}$ 
and $B=70\;{\rm MeV\,fm^{-3}}$ \cite{AFO,HZS,Var1,Var2,Var3}, $B=109\;{\rm MeV\,fm^{-3}}$\,\cite{Koh}, 
and $B=208\;{\rm MeV\,fm^{-3}}$ and $B=508\;{\rm MeV\,fm^{-3}}$\,\cite{Ber}. 
 Along the $V_1=0$ curve and at $\Delta_2M=0$ is our calculated value of 
$\Delta|\varepsilon_{\rm vac}|$=$B=170\;{\rm MeV\,fm^{-3}}$ corresponding to 
$G_2=0.00927\,{\rm GeV}^4$.

{\bf Fig. 5 -}  Comparison between mass defects taking $m_A=m_n$ and $m_A=m_0$.

{\bf Fig. 6 -}  Panel (a): SQM stability windows at zero temperature and pressure, bounded by 
the $E/A=0.9304\,{\rm GeV}$ (solid) contour of $^{56}F_\ex$ and its vertical 
(dashed) line, for different choices of $V_1$ (in GeV units) labeling each contour 
(cf. panel (a) of \;Fig.\;3 in Ref.\;\cite{Fla2}).
 Along the horizontal straight (dashed) line, the width of the stability window is the 
distance from the vertical line to the corresponding $E/A$ contour at a given $V_1$. 
 Panel (b): The relative SQM stability as function of $V_1$ estimated according to 
three different ways. 
 For each value of $V_1$:
 (solid line) the width of the stability window  at $m_s=0.15\,{\rm GeV}$ divided by 
the one at $m_s=0$ (at the same $V_1$); 
 (short dashed line) the width of the stability window  at $m_s=0.15\,{\rm GeV}$ divided by 
the one at $V_1=0$ (at the same $m_s=0.15\,{\rm GeV}$); 
 (long dashed line) the width of the stability window at $m_s=0.15\,{\rm GeV}$ divided 
by the largest one (at $V_1=0$ and $m_s=0$)\,.

\newpage


\newpage


\newpage


\newpage


\newpage


\newpage


\end{document}